\documentstyle[astro-cite,additional,psfig_laa]{laa}
\begin{document}
\thesaurus{Interstellar Matter (02.18.7, 09.13.2., 09.19.1, 10.03.1,
11.14.1, 13.19.3)}
\title{Molecular Gas in the Galactic Center Region}
\subtitle{II. Gas mass and ${\cal N}_{\bf H_2}$/$I_{\bf ^{12}CO}$
conversion based on a C$^{18}$O($J = 1 \to 0$) Survey}
\author{
G.~Dahmen\inst{1,3}, 
S.~H\"uttemeister\inst{2,4,1}, 
T.L.~Wilson\inst{1,5},
R.~Mauersberger\inst{6,1}
}
\offprints{S.\ H\"uttemeister, Radioastronomisches Institut,
           Universit\"at Bonn}
\institute{
 Max-Planck-Institut f\"ur Radioastronomie, 
 Auf dem H\"ugel 69, 53121 Bonn, Germany 
\and
 Harvard-Smithsonian Center for Astrophysics, 
 60 Garden Street, Cambridge, MA 02138, U.S.A. 
\and
 Physics Department, 
 Queen Mary \& Westfield College, 
 University of London, 
 Mile End Road, London E1 4NS, England 
\and
 Radioastronomisches Institut, Universit\"at Bonn, 
 Auf dem H\"ugel 71, 53121 Bonn, Germany 
\and
 Submillimeter Telescope Observatory, The University of
 Arizona, Tucson, AZ 85721, U.S.A. 
\and
 Steward Observatory, The University of Arizona, 
 Tucson, AZ 85721, U.S.A.
}
\date{Received / Accepted}
\maketitle
\markboth{G.\ Dahmen et al.: Molecular Gas in the Galactic Center Region, 
          II. Gas mass and ${\cal N}_{\rm H_2}$/$I_{\rm ^{12}CO}$
              conversion~\dots}{ }

\begin{abstract}
The large scale structure and physics of molecular gas in the Galactic
center region is discussed based on the detailed analysis of a
9\arcmin{} resolution survey of the Galactic center region in the $J =
1 \to 0$ line of C$^{18}$O. Emphasis is placed on the comparison with
\mbox{$^{12}$CO(1--0)} data.  The line shapes of \mbox{C$^{18}$O(1--0)}
and \mbox{$^{12}$CO(1--0)} differ significantly.  The ratio of the
intensities of the two isotopomers in the Galactic center region is
generally higher than the value of $\sim$\,15 expected from the
``Standard Conversion Factor'' (SCF) of $^{12}$CO integrated line
intensity to H$_2$ column density.  In the 9\arcmin{}-beam, this ratio
is in the range from 30 to 200, mostly $\sim$\,60 to 80. From LVG
calculations, we estimate that the large scale \mbox{$^{12}$CO(1--0)}
emission in the Galactic center region is of moderate ($\tau \grsim 1$)
or low optical depth ($\tau < 1$). Higher optical depths ($\tau \ge
10$) are restricted to very limited regions such as Sgr\,B2.  In
addition, we estimate H$_2$ densities and kinetic temperatures for
different ranges of intensity ratios.  A considerable amount of
molecular mass is in a widespread molecular gas component with low
densities and high kinetic temperatures. From our C$^{18}$O
measurements and from results based on dust measurements, the total
molecular mass is found to be $(3^{+2}_{-1}) \cdot 10^7 \, {\rm
M_{\odot}}$. We show that the SCF is {\it not\/} valid toward the
Galactic bulge. It overestimates the H$_2$ column density by an order
of magnitude because the assumptions required for this factor of
optically thick $^{12}$CO emission and virialization of the molecular
clouds are not fulfilled for a significant fraction of the molecular
gas. Therefore, also one cannot apply a modified conversion factor to
the Galactic center region since the ${\cal N}_{\rm H_2}/I_{\rm ^{12}CO}$
is highly variable and cannot be represented by a universal constant. 
Results from external galaxies indicate that the $^{12}$CO emission is
generally not a suitable tracer of H$_2$ masses in galactic bulges.
\end{abstract}


\begin{figure*}
\psfig{figure=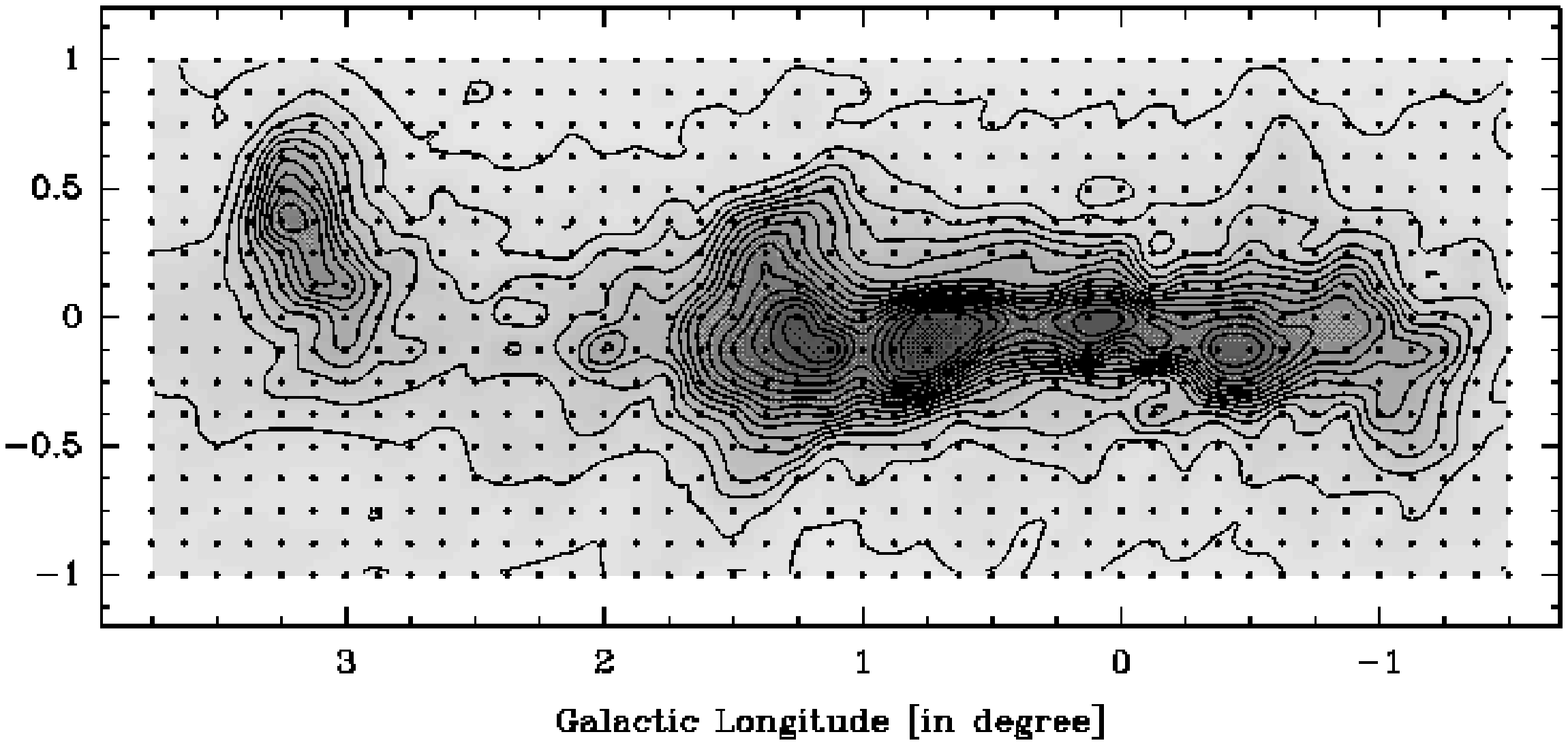,bbllx=430pt,bblly=25pt,bburx=115pt,bbury=770pt,angle=-90,width=\textwidth}
\caption{The integrated intensity of the Galactic center region in
\mbox{C$^{18}$O(1--0)} (from Paper~I). The velocity over which the
intensity (in \Tmb{}) is integrated ranges from $-$225.0 to
+225.0~\kms.  The solid contour levels range from 3.9 to 28.05 in steps
of 3.45~K\,\kms{} where the lowest level is the 3$\sigma$-value. The
dashed contour is at 2.6~K\,\kms{} which is the 2$\sigma$-value. The
circle in the lower left corner of the plot indicates the beam size of
9\farcm{}2.} \label{C18OContAll}
\end{figure*} 

\section{Introduction} \label{Introduct}

Molecular clouds close to the center of our Galaxy are distinctly
different from molecular material in the disk. Being dense, hot and
highly turbulent \citep[see, e.g., ]{Hut+93b}, they constitute a third
type of molecular environment besides cold and dense dark clouds and
the warm and less dense surroundings of H\,II regions found in the
Galactic disk. Within an area of a few hundred parsecs, these clouds
concentrate $\sim$\,10\% of the total neutral gas mass of the Galaxy
\citep[see, e.g., ]{Gus89}.  However, the properties of the molecular
gas in this central region are less understood than other regions of
the Milky Way.

Particularly important is the determination of the mass of molecular
clouds, 70\% of which is made up by molecular hydrogen, H$_2$.
However, H$_2$ lacks a permanent dipole moment and, hence, allowed
rotational dipole transitions. Thus, the presence and properties of
H$_2$ in molecular clouds must be deduced from ``tracer molecules''
with permanent dipole moments.  Such analyses must account for the
excitation effects and the abundance of the molecules which produce the
observable lines.

The most abundant molecule in the ISM with permanent dipole moment is
carbon monoxide \citep[see, e.g., ]{Irv+85}. Because of the low
critical density, $n^*$, of its $J = 1 \to 0$ line of about 740~\pccm,
CO traces even the low-density molecular phase of the ISM.  However,
the emission of the main isotopomer $^{12}$C$^{16}$O is in most cases
optically thick making it difficult to determine the volume and column
densities from its emission lines.  A comparison of
$^{12}$CO\footnote{Following typical astronomical conventions, the
atomic weight of the main isotopes will not be given explicitly, e.g.,
$^{13}$CO and C$^{18}$O. However, for clarity we denote CO as
$^{12}$CO.} with $^{13}$CO maps from large scale surveys in the $J = 1
\to 0$ line in our Galaxy and external galaxies has shown that the line
shapes and intensity ratios along different lines of sight are
remarkably similar for both isotopomers, and the ratio of the CO
isotopomer intensities varies remarkably little for different regions
in the Galaxy \citep{Dic75,YoSc82,San+84}. Because the weaker
$^{13}$CO(1--0) line was usually assumed to be optically thin and,
therefore, to trace the column density of H$_2$ directly, it was
concluded that the $^{12}$CO luminosity, $L_{\rm ^{12}CO(1-0)}$, also
measures mass, even though this line is optically thick \citep[see, e.g.,
]{Dam93}. This empirical $^{12}$CO-mass relation was confirmed, among
other things, by comparison with $\gamma$-ray fluxes emitted when cosmic
rays interact with H$_2$ molecules \citep{Blo+86} and by comparison with
virial masses of Galactic and extragalactic clouds determined from their
diameters and velocity dispersion \citep{Sol+87}.

At first glance, it seems paradoxical that an optically thick
transition can give information about the total column density of
H$_2$. The paradox can be solved if one assumes that there are several
clouds within a beam area, the emission of which does not overlap in
space and velocity (i.e.\ ``CO counts clouds''). Also, if virialized,
more massive clouds have more turbulence and, hence, a larger linewidth.
Finally, the denser the gas the higher is the excitation of the gas. All
these effects combined can approximately explain the observed relation
\citep[see, e.g., ]{Dic+86,Tay+93}. A number of investigations have been
used to calibrate $L_{\rm ^{12}CO(1-0)}$ to the total molecular mass,
both for the disk of the Milky Way and for external galaxies. This has
led to establishing the ``Standard Conversion Factor'' (hereafter
SCF), a.k.a.\ X, by \cite{Str+88} which relates the integrated intensity
of \mbox{$^{12}$CO(1--0)} to the H$_2$ column density:
\begin{equation} \label{Eq-H2-from-CO} 
\bar{\cal N}_{\rm H_2} = 2.3 \cdot 10^{20} \, \frac{\rm cm^{-2}}
                      {\rm K\,km\,s^{-1}} \, \int 
                      T_{\rm MB}^{\rm ^{12}CO(1-0)} \, {\rm d}v
\end{equation}

\begin{figure*}
\psfig{figure=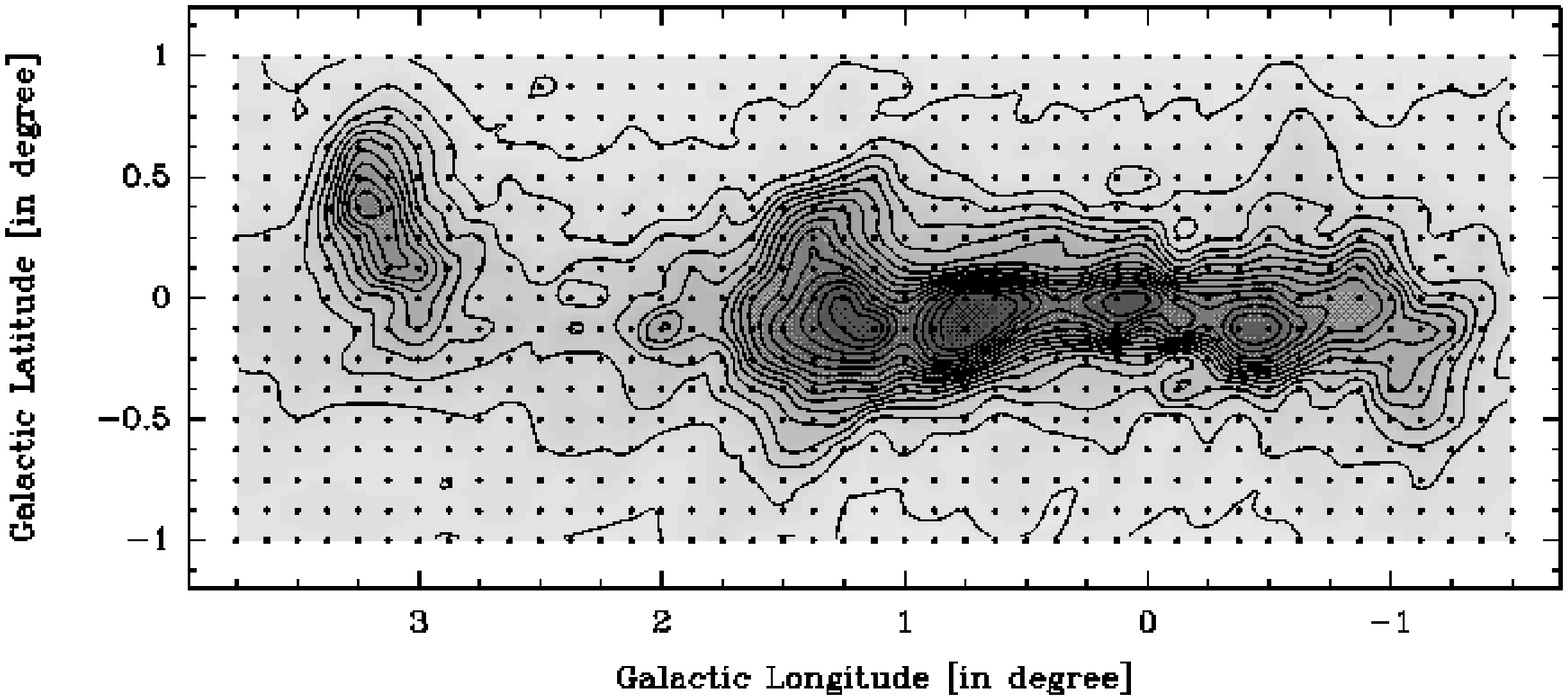,bbllx=460pt,bblly=25pt,bburx=115pt,bbury=785pt,angle=-90,width=\textwidth}
\caption{The integrated intensity of the Galactic center region in
\mbox{$^{12}$CO(1--0)} (data taken from \protect\citeo{Bit87}). The
velocity the intensity (in \Tmb{}) is integrated over ranges from
$-$225.0 to +225.0~\kms. The plotted section ranges from $l=-1$\fdg{}5
to +3\fdg{}75 and from $b=-1$\fdg{}0 to +1\fdg{}0.  The solid contour
levels range from 50.0 to 1750.0 in steps of 100.0~K\,\kms. The minimum
in the contour map is well above the 3$\sigma$-value of about
9.6~K\,\kms. The circle in the lower left corner of the plot indicates
the beam size of 8\farcm{}8.} \label{12COContAllBit}
\end{figure*} 

We presented a large scale survey of the Galactic center region in
the $J = 1 \to 0$ line of C$^{18}$O obtained with the 1.2\,m Southern
Millimeter-Wave Telescope (SMWT) at the Cerro Tololo Interamerican
Observatory (CTIO) near La Serena, Chile, in \cite{Dah+97a}, hereafter
Paper~I. Here, we carry out a more detailed analysis emphasizing the
comparison to \mbox{$^{12}$CO(1--0)} data obtained with the same
telescope. A summary of the conclusions was given in \cite{Dah+96}.  In
Section~\ref{Data_Analysis}, spectral line shapes toward selected
positions are analyzed and ratios of integrated intensities of
\CO{12}/\CO{18} are determined.  In addition, we estimate optical
depths, volume densities, and kinetic temperatures from Large Velocity
Gradient (LVG) radiative transfer calculations.
From these results, the H$_2$ mass of the Galactic bulge is estimated.
In Section~\ref{Discuss}, we compare our molecular mass to results
based on dust measurements and give a {\em weighted best estimate\/}
for the total molecular mass in the Galactic bulge.  The consequences,
e.g.\ for other galactic nuclei, are discussed.  Finally, in
Section~\ref{Conclu}, we present the conclusions drawn from our
analysis.

Throughout this paper we adopt a distance to the Galactic center of
8.5~kpc.

\section{Data and Analysis} \label{Data_Analysis}

The data of the \mbox{C$^{18}$O(1--0)} Galactic Center Survey were
taken from August 1993 to August 1994 with the 1.2\,m SMWT at CTIO. The
survey covers the region $-$1\fdg{}$05 \leq l \leq +3$\fdg{}6 and
$-$0\fdg{}$9 \leq b \leq +0$\fdg{}75 and has a spacing of 9\arcmin{}
which is the FWHP beamwidth (half sampling). A detailed description of
these data and the observations is given in Paper~I. In
Fig.~\ref{C18OContAll}, we show the distribution of the integrated
intensity of \mbox{C$^{18}$O(1--0)}, based on this survey.

For comparison, we used the \mbox{$^{12}$CO(1--0)} Galactic Center
Survey, observed with the same telescope from February to November 1984
\citep{Bit87,Bit+97}. This survey covers a larger area than the region
observed in \CO{18} and has a sampling of 7\farcm{}5. These data were
made available to us in digital form, allowing a detailed analysis.  A
comparison of \mbox{$^{12}$CO(1--0)} data taken recently with the
1.2\,m SMWT (see Paper~I) with \citen{Bit87}'s observations shows that
the data sets are compatible (see \citeo{Dah95} for the full account).
The scaling factor to obtain \Tmb{} from \citen{Bit87}'s \Tastar{}
scale is 1.11 (see \citeo{Dah95} for a discussion).  In
Fig.~\ref{12COContAllBit}, we show the distribution of the integrated
intensity of \mbox{$^{12}$CO(1--0)}, based on \citen{Bit87}'s data.

As shown in Paper~I, the spectra and maps of the \mbox{C$^{18}$O(1--0)}
survey demonstrate the great differences in the distribution of the
optically thin \mbox{C$^{18}$O(1--0)} emission and the
\mbox{$^{12}$CO(1--0)} emission, which is generally assumed to be
optically thick. In addition, the $^{12}$CO emission is much more
widespread than the C$^{18}$O emission whereas the C$^{18}$O emission
shows much more contrast.

\begin{figure*}
\psfig{figure=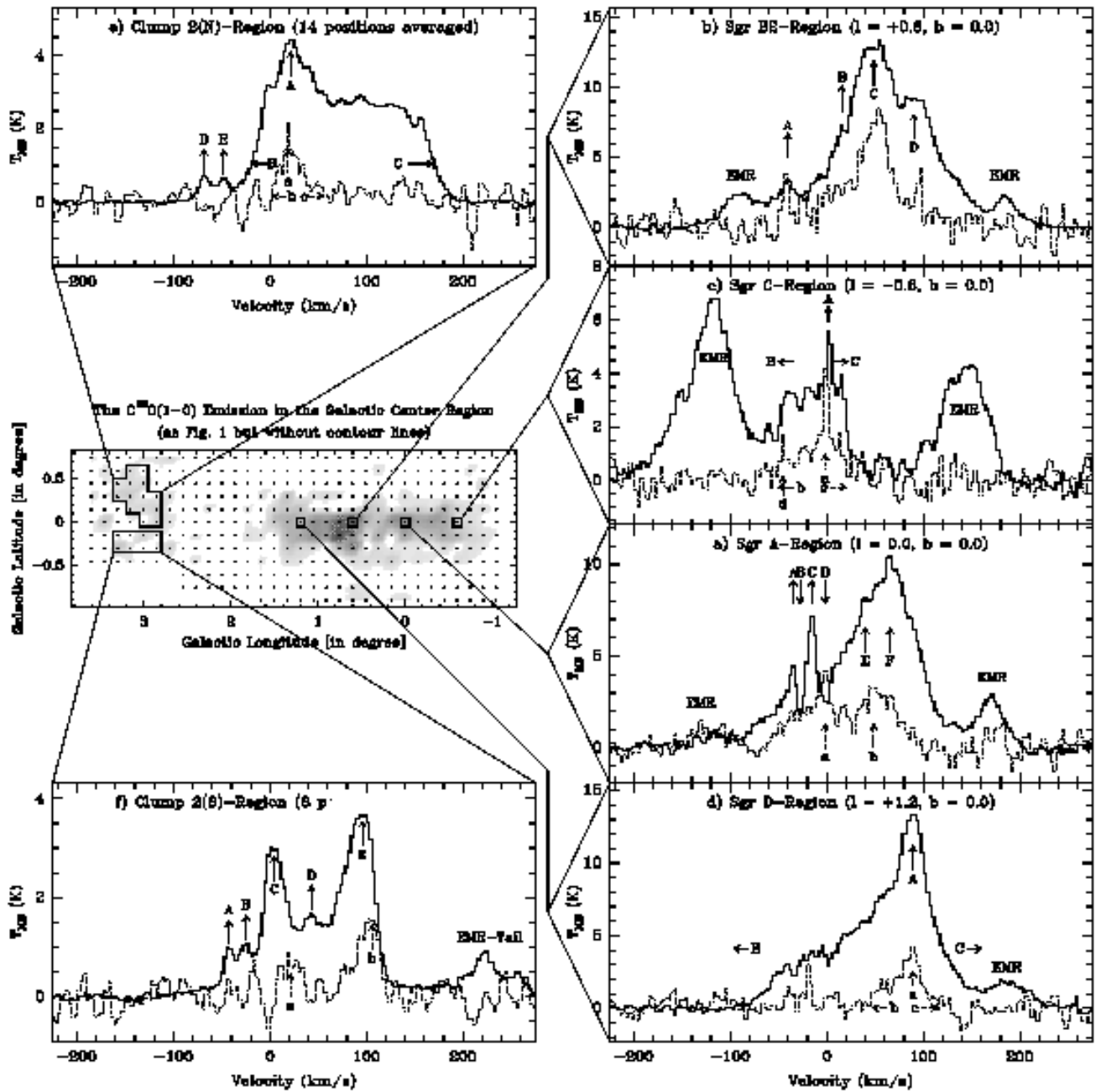,%
bbllx=555pt,bblly=150pt,bburx=45pt,bbury=680pt,%
angle=-90,width=\textwidth}
\caption{CO isotopomer spectra toward characteristic emission centers.
In all plots, the solid histogram is the spectrum of the
\mbox{$^{12}$CO(1--0)} transition, while the dashed histogram is the
spectrum of the \mbox{C$^{18}$O(1--0)} transition, multiplied by 20 for
better comparability. Both were observed with the same SMWT setup
during the C$^{18}$O Galactic Center Survey. Capital letters always
mark features in \CO{12}, lower case letters features in \CO{18}.
Upward pointing arrows indicate emission features, downward pointing
arrows dips in the spectra.  {\bf a)} Position ($l = 0$\fdg{}0, $b =
0$\fdg{}0) near Sgr\,A. {\bf b)} Position ($l = +0$\fdg{}6, $b =
0$\fdg{}0) near Sgr\,B2. {\bf c)} Position ($l = -0$\fdg{}6, $b =
0$\fdg{}0) near Sgr\,C. {\bf d)} Position ($l = +1$\fdg{}2, $b =
0$\fdg{}0) near Sgr\,D \protect\citep[$l=1$\fdg{}5-complex of
]{Bal+88}.  {\bf e)} Northern Clump~2 region (14 positions averaged as
indicated).  {\bf f)} Southern Clump~2 region (8 positions averaged as
indicated).} \label{VerglSpectra}
\end{figure*}

\begin{figure*}
\psfig{figure=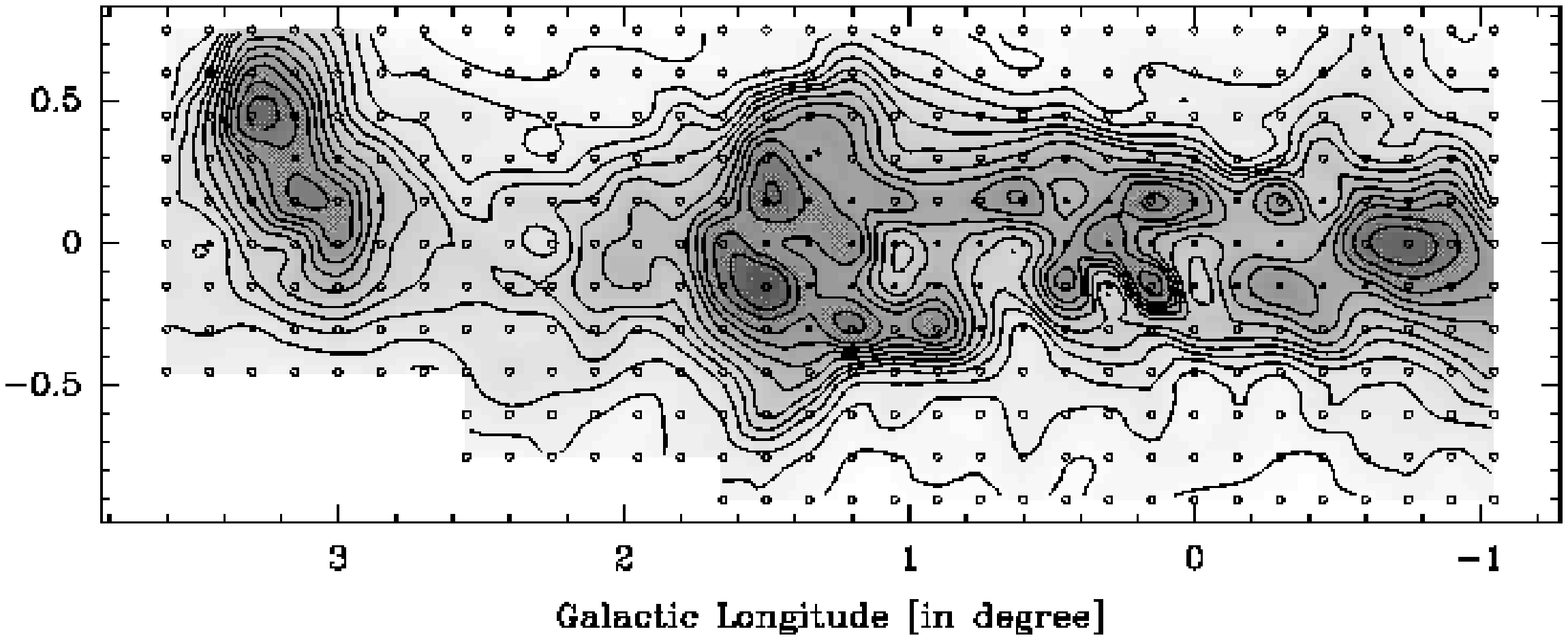,bbllx=430pt,%
bblly=25pt,bburx=170pt,bbury=730pt,angle=-90,width=\textwidth}
\caption{The integrated intensity ratio of \mbox{$^{12}$CO(1--0)} to
\mbox{C$^{18}$O(1--0)} emission in the Galactic center region. The
intensities are integrated over a velocity range from $-$225.0 to
+225.0~\kms. The solid contour levels range from 10.0 to 60.0 in steps
of 10.0, from 60.0 to 120.0 in steps of 15.0, and from 120.0 to 180.0
in steps of 20.0. The circle in the lower left corner of the plot
indicates the averaged beam size of 9\arcmin. The ratio was calculated
with an 3$\sigma$-threshold, thus, if the integrated intensity of
C$^{18}$O was below 3$\sigma$ r.m.s.\ the 3$\sigma$ value was taken
instead for the calculation of the ratio. Filled points indicate
positions where C$^{18}$O is above this threshold. Open points indicate
positions where C$^{18}$O is below this threshold, thus, positions
where the calculated ratio is a lower limit.} \label{RatioContAll}
\end{figure*} 

\subsection{Spectral Line Shapes toward Selected Positions} 

In Fig.~\ref{VerglSpectra}, we show spectra of both CO isotopomers
toward characteristic emission centers, four examples of single
positions toward the Galactic bulge and two of Clump~2 \citep{Ban77}
averaged over several positions to improve the signal-to-noise ratio of
the weak C$^{18}$O emission. The \CO{12} spectra used were observed in
March 1994 and August 1994 with the same SMWT setup as the \CO{18}
Galactic Center Survey, as presented in Paper~I. Throughout the Galaxy,
the line shapes and intensity ratios of the optically thick $^{12}$CO
and the optically thin $^{13}$CO transition along different lines of
sight in the Galactic disk are remarkably similar, as mentioned in
Section~\ref{Introduct}. This is also what should be the case if the
SCF (Eq.~\ref{Eq-H2-from-CO}) applies. However, in
Appendix~\ref{LineShape} where we discuss the presented spectra in
detail, we show that this is in many cases (e.g.\ plots (a), (c), (e)
and (f) of Fig.~\ref{VerglSpectra}) not true for the Galactic center
\CO{12} and \CO{18}.  Therefore, one or more of the requirements of
this factor are not fulfilled (see also Section~\ref{12COConvFac}).
This alone is proof that the $J = 1 \to 0$ lines of $^{12}$CO and
C$^{18}$O {\it cannot\/} both trace the total H$_2$ mass in the
Galactic center region, as it is commonly assumed for disk clouds. We
also show that even toward positions where, at first, it appears that
the SCF might be valid (e.g.\ plots (b) and (d) of
Fig.~\ref{VerglSpectra}) this is not the case. On the contrary, based
on an LTE analysis we show that the data suggest moderate optical
depths ($\tau \grsim 1$) for most of the $^{12}$CO emission, even
toward very intense peaks, rather than very high optical depths ($\tau
\gg 1$) which is one of the essential requirements that the SCF is
applicable. From these results, we conclude that in the Galactic center
region:
\begin{enumerate}
\item the SCF for the determination of the H$_2$ column density is
      not applicable;
\item the assumption of large optical depth for the 
      \mbox{$^{12}$CO(1--0)} emission is not valid. 
\end{enumerate}

\subsection{CO Integrated Intensity Ratios} \label{COIntRat} 

From a comparison of Figs.~\ref{C18OContAll} and \ref{12COContAllBit},
it is obvious that there are great differences in the spatial
distribution of the integrated intensities of \mbox{C$^{18}$O(1--0)}
and \mbox{$^{12}$CO(1--0)}. These differences become even larger if one
compares the channel maps of integrated intensities (see
Figs.~\ref{C18OContChan} and \ref{12COContChan} in
\pageref{12COContChan}) or the $lv$-plots (see Figs.~\ref{C18OLVplot}
and \ref{12COLVplot} in \pageref{12COLVplot}).  The C$^{18}$O emission
has a distinctly larger contrast compared to the rather smooth
$^{12}$CO emission. In this section, we investigate the variation in
the intensity ratio of $^{12}$CO/C$^{18}$O, first integrated over the
total velocity extent of CO emission, and then over velocity intervals
of 50~\kms{} width.

\begin{figure*}
\psfig{figure=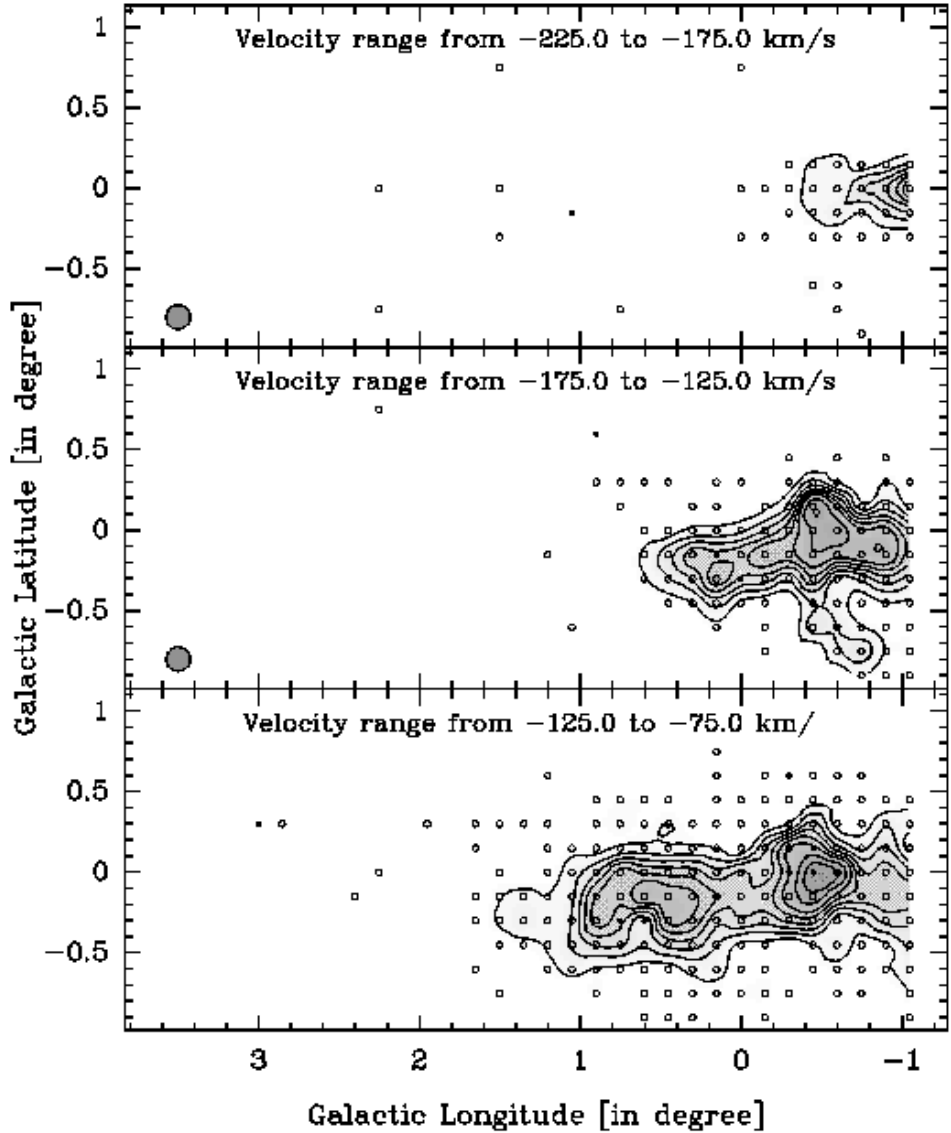,%
bbllx=0pt,bblly=90pt,bburx=515pt,bbury=750pt,height=16cm}
\caption{The intensity ratio of \mbox{$^{12}$CO(1--0)} to
\mbox{C$^{18}$O(1--0)} emission in the Galactic center region
integrated over velocity intervals of 50~\kms{} width. The solid
contour levels range from 10.0 to 60.0 in steps of 10.0, from 60.0 to
120.0 in steps of 15.0, and from 120.0 to 200.0 in steps of 20.0. The
circle in the lower left corner of the plots indicates the averaged
beam size of 9\arcmin. The ratio was calculated with an
3$\sigma$-threshold, thus, if the integrated intensity of either
C$^{18}$O or $^{12}$CO was below 3$\sigma$ r.m.s.\ the 3$\sigma$ value
was taken instead for the calculation of the ratio. Filled points
indicate positions where both C$^{18}$O and $^{12}$CO are above the
threshold. Open points indicate positions where C$^{18}$O is below the
threshold but where $^{12}$CO emission is seen, thus, where the
calculated ratio is a lower limit. Where both C$^{18}$O and $^{12}$CO
are below the threshold, no statement about the $^{12}$CO/C$^{18}$O
ratio can be made, so these positions are not indicated.  {\bf a)} At
the top the integrated intensity ratio of the velocity range from
$-$225.0 to $-$175.0~\kms{} is plotted, in the middle panel the
velocity ranges from $-$175.0 to $-$125.0~\kms, and at the bottom the
velocity range from $-$125 to $-$75.0~\kms{} is shown.}
\label{RatioContChan}
\end{figure*}

\begin{figure*}
\psfig{figure=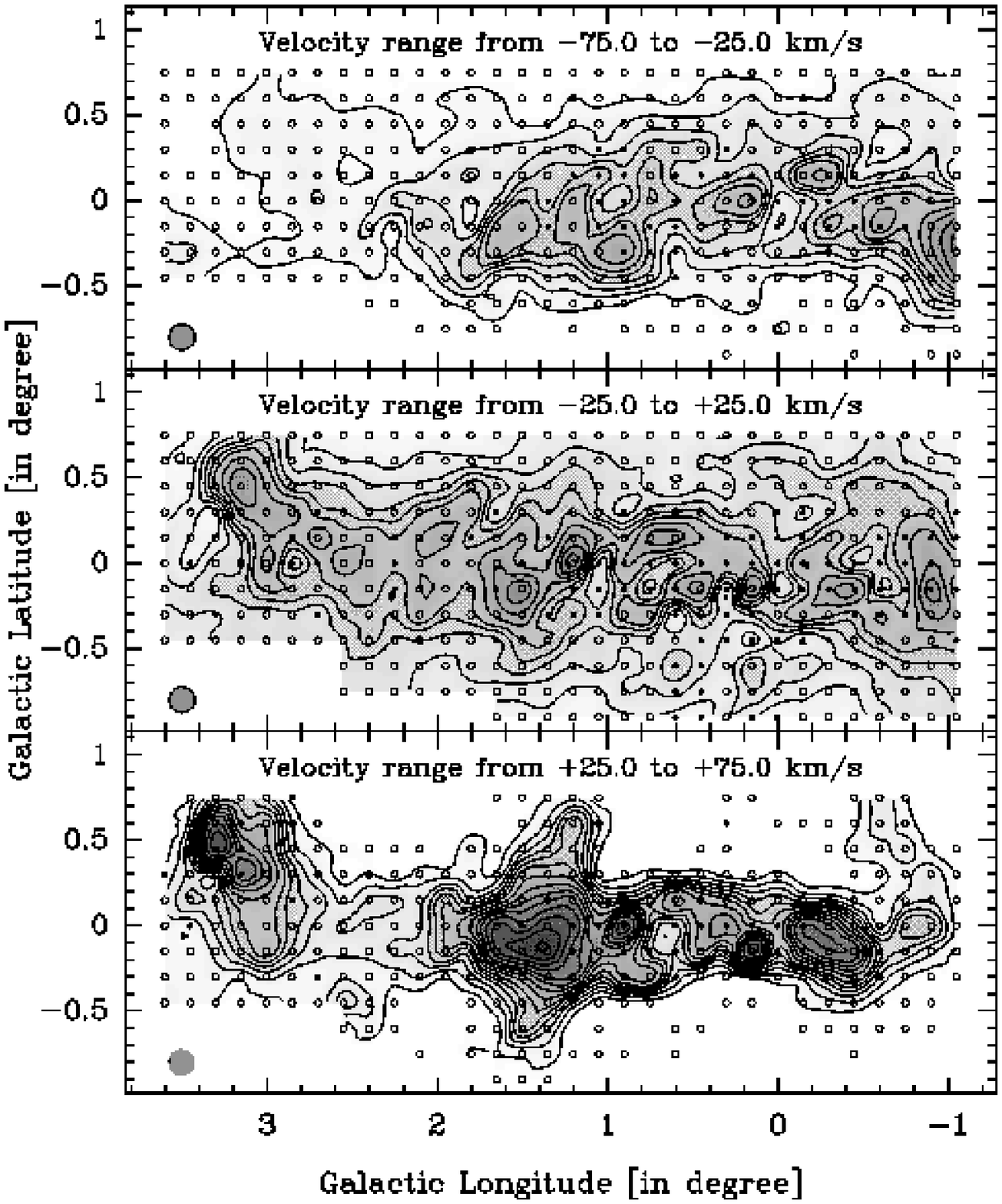,%
bbllx=0pt,bblly=90pt,bburx=515pt,bbury=750pt,height=16cm}
\ccaption{{\bf b)} At the top the integrated intensity ratio of the
velocity range from $-$75.0 to $-$25~\kms{} is plotted, in the middle
panel the velocity ranges from $-$25 to +25~\kms, and at the bottom the
velocity range from +25 to +75.0~\kms{} is shown.}
\end{figure*}

\begin{figure*}
\psfig{figure=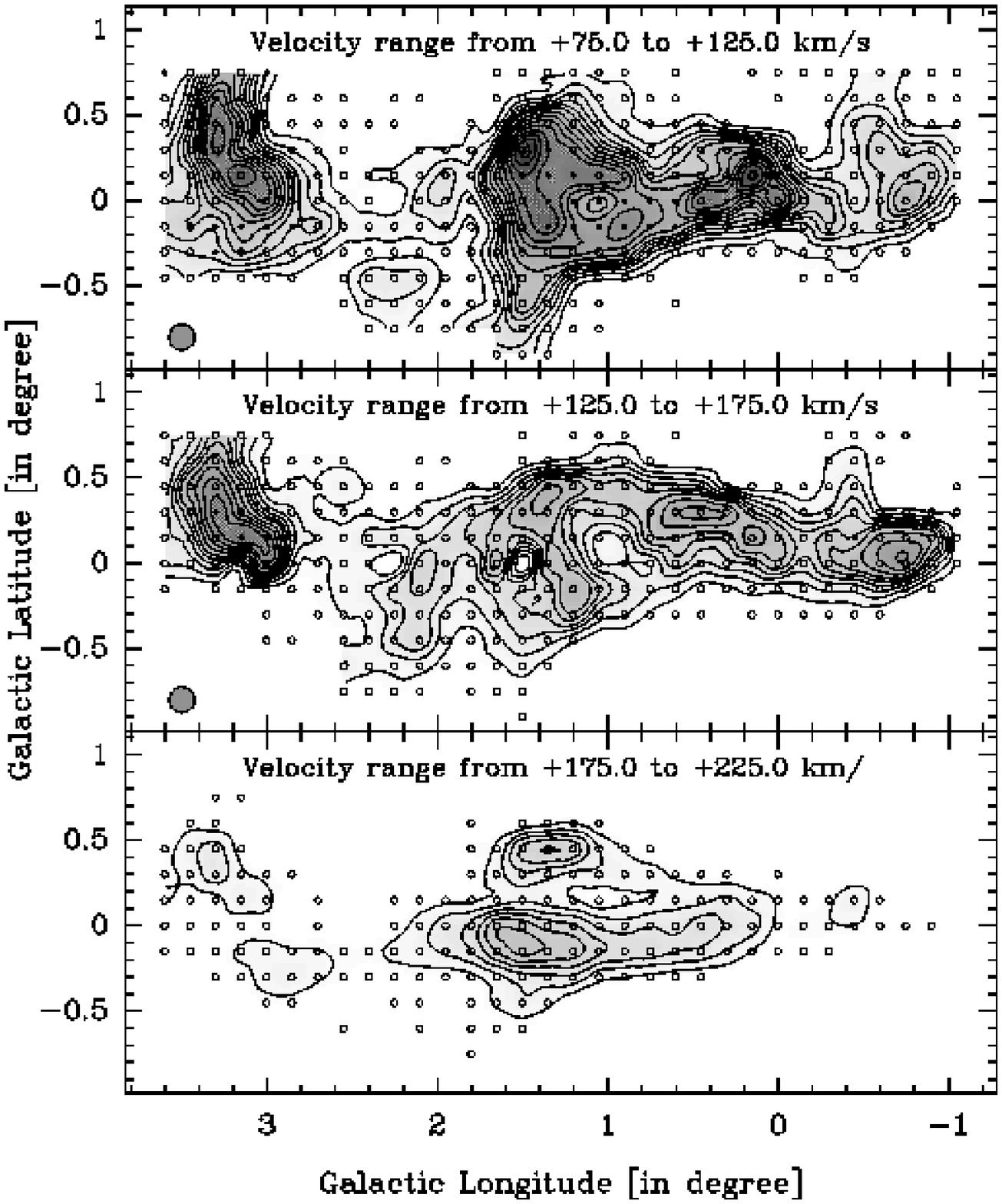,%
bbllx=0pt,bblly=90pt,bburx=515pt,bbury=750pt,height=16cm}
\ccaption{{\bf c)} At the top the integrated intensity ratio of the
velocity range from +75.0 to +125~\kms{} is plotted, in the middle
panel the velocity ranges from +125 to +175~\kms, and at the bottom the
velocity range from +175 to +225.0~\kms{} is shown.}
\end{figure*}

For this purpose, the \mbox{$^{12}$CO(1--0)} data of \cite{Bit87} were
resampled to the slightly lower velocity and spatial resolution of the
C$^{18}$O spectra. Then, the integrated intensity ratios were
determined by calculating the integrated intensities of $^{12}$CO and
C$^{18}$O for the chosen velocity interval respectively and dividing
the result for $^{12}$CO by the result for C$^{18}$O.  The ratios of
the integrated intensities were calculated with a threshold set to 3
times the r.m.s.\ noise (see below for its value). Thus, if the
integrated intensity in $^{12}$CO or in C$^{18}$O was below this
3$\sigma$-limit, the 3$\sigma$-value was taken instead.  Because the
$^{12}$CO emission is much more intense than the C$^{18}$O emission,
there are areas where $^{12}$CO is present but C$^{18}$O is below the
detection limit. In these cases, the determined ratio of the integrated
intensities is a {\it lower limit\/} instead of the true value.
Similarly, calculated ratios of the integrated intensities are {\it
never\/} upper limits, hence, the true ratio is {\it never\/} lower
than calculated (within the noise scatter). The lowest possible value
is about 2.5, the ratio of the 3$\sigma$-values (which is equivalent to
no information on the ratio).

For $^{12}$CO, we used as the 3$\sigma$-threshold the value which
should statistically be achieved after integrating over the desired
velocity range (and which is $3\cdot \sqrt{N_{ch}}\ v_{ch}\
r.m.s._{ch}$; see also Eq.~\ref{AreaRMS} in \pageref{AreaRMS}). This
was 14.2~K\,\kms{} in case of intensities integrated over of the total
velocity range and 4.9~K\,\kms{} in case of intensities integrated over
velocity intervals of 50~\kms{} width.

In case of C$^{18}$O, however, it turned out that this straight-forward
method is not applicable.  Because of systematic effects the noise in
our data improves more slowly than the square root of the integration
time.  Thus, the true r.m.s.\ of the integrated intensities cannot be
found as easily as in case of $^{12}$CO. It is, however, essential for
our analysis that we do not use a threshold which is too low because in
this case we might overestimate the $^{12}$CO/C$^{18}$O ratio instead
of using a valid lower limit. Because our conclusions rely critically
on high $^{12}$CO/C$^{18}$O ratios, we must be certain that we never
overestimate this ratio. Therefore, we have to determine the r.m.s.\ of
the integrated intensities empirically. We carried this out by
calculating the r.m.s\ deviation of the intensities integrated over
velocity intervals of 50~\kms{} width for the positions from $l =
-1$\fdg{}05 to +2\fdg{}1 at $b = +0$\fdg{}75.  These 22 positions are
ensured to be emission-free in C$^{18}$O over the complete velocity
interval. We find an r.m.s.\ of 0.65~K\,\kms. Hence, we choose as
3$\sigma$-threshold for C$^{18}$O intensities integrated over velocity
intervals of 50~\kms{} a value of 2.0~K\,\kms.  Similarly, we find the
3$\sigma$-threshold for C$^{18}$O intensities integrated over of the
total velocity range to be 5.9~K\,\kms.

In Fig.~\ref{RatioContAll}, we plot the ratio of the
\mbox{$^{12}$CO(1--0)} to \mbox{C$^{18}$O(1--0)} intensity in the
Galactic center region, integrated over the total velocity interval
from $-$225.0 to +225.0~\kms. The ratio of the integrated intensities
shows a large variation from 30 to 190. The most prominent maxima are
the Sgr\,D region \citep[$l=1$\fdg{}5-complex of ]{Bal+88} with ratios
up to 190 at ($l = +1$\fdg{}5, $b = -0$\fdg{}15), the region somewhat
west of Sgr\,C with ratios up to 175 at ($l = -0$\fdg{}75, $b =
0$\fdg{}0), and Clump~2 with ratios up to 165 at ($l = +3$\fdg{}3, $b =
+0$\fdg{}45). On the other hand, true minima (not lower limits),
valley-likely surrounded by higher ratios, are visible toward the
regions of Sgr\,A and Sgr\,B2 and toward the region at ($l =
+1$\fdg{}05, $b = 0$\fdg{}0).

High integrated intensity ratios tend to coincide with strong CO
emission. Since high ratios are correlated with low optical depths in
both \mbox{$^{12}$CO(1--0)} and \mbox{C$^{18}$O(1--0)}
(Eq.~\ref{Eq-tau12-RatInt}), one might expect these maxima to be
correlated with weak CO emission. However, at positions where the
$^{12}$CO emission is weak, the strength of the C$^{18}$O emission
might be {\it far below\/} the 3$\sigma$-threshold, thus, the
determined {\it lower limit\/} for the integrated intensity ratio might
underestimate the true ratio {\it by far\/}.

In Fig.~\ref{RatioContChan}, we show the ratios of $^{12}$CO(1--0) to
C$^{18}$O(1--0) intensities integrated over velocity intervals of
50~\kms{} width. While already the ratio of the intensities integrated
over the total velocity interval shows a large variation and a lot of
structure this is even more so in the 50~km\,s$^{-1}$-interval maps. In
addition, the ratio in the channel maps reaches values up to $>$\,200.

In Appendix~\ref{FeatFigRatCoCh}, we give a detailed description of the
features visible. This analysis shows that the integrated intensity
ratio of $^{12}$CO/C$^{18}$O in the Galactic center region is generally
higher than the value of $\sim$\,15 expected from the ``standard''
conversion factors. Wherever C$^{18}$O is above the detection limit,
this ratio is at least of order 40, mostly of order of 60 to 80, in
several areas of order 90 to 120, and toward a few positions even up to
$>$\,200. Note that these values are an average over an area of
9\arcmin{} diameter. Therefore, we conclude that high
$^{12}$CO/C$^{18}$O ratios are typical of the molecular gas in the
Galactic center region. Ratios as low as expected from the ``standard''
conversion factors are rare and restricted to rather small areas in the
Galactic center. For example, from measurements with the IRAM 30\,m
telescope (23\arcsec{} resolution) \cite{Mau+89} found a
$^{12}$CO/C$^{18}$O ratio of 13 toward Sgr\,B2. Thus, if the beam is
small and the observation carried out toward the peak of an exceptional
region, the ratio can be in agreement with the standard conversion.
However, averaged over the 9\arcmin-beam of the 1.2\,m SMWT, the ratio
is $\losim$\,40 (see Appendix~\ref{FeatFigRatCoCh}).  Thus, the lower
ratio is found only in a very limited area close to the peak of
Sgr\,B2.

\subsection{LVG Calculations on the CO Isotopomer Ratio} \label{LVGCalc} 

In this section, we carry out radiative transfer calculations for the 
excitation of the $J = 1 \to 0$ transition of $^{12}$CO and C$^{18}$O 
using the Large Velocity Gradient (LVG) approximation with collision
rates taken from \cite{GrCh78} to explore non-LTE conditions (see,
e.g., \cite{dJo+75} or \cite[chapter 14.10, ]{RoWi96} for a full
description of radiative transfer in the LVG approximation).  The LVG
radiative transfer program used was kindly provided by C.\ Henkel; see
\cite{Hen80} for a full description of the program.

\begin{figure}
\psfig{figure=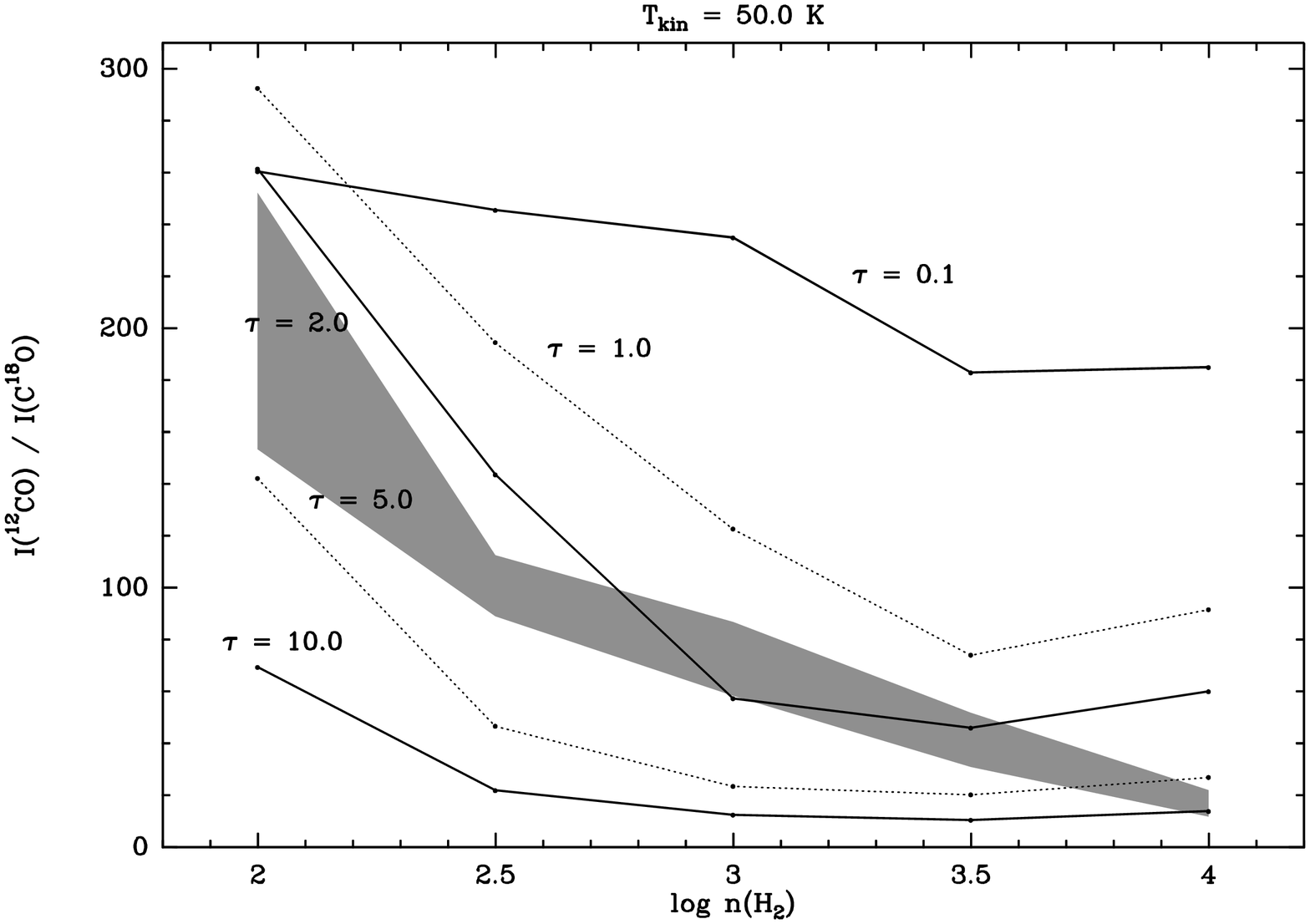,angle=-90,width=8.75cm}
\psfig{figure=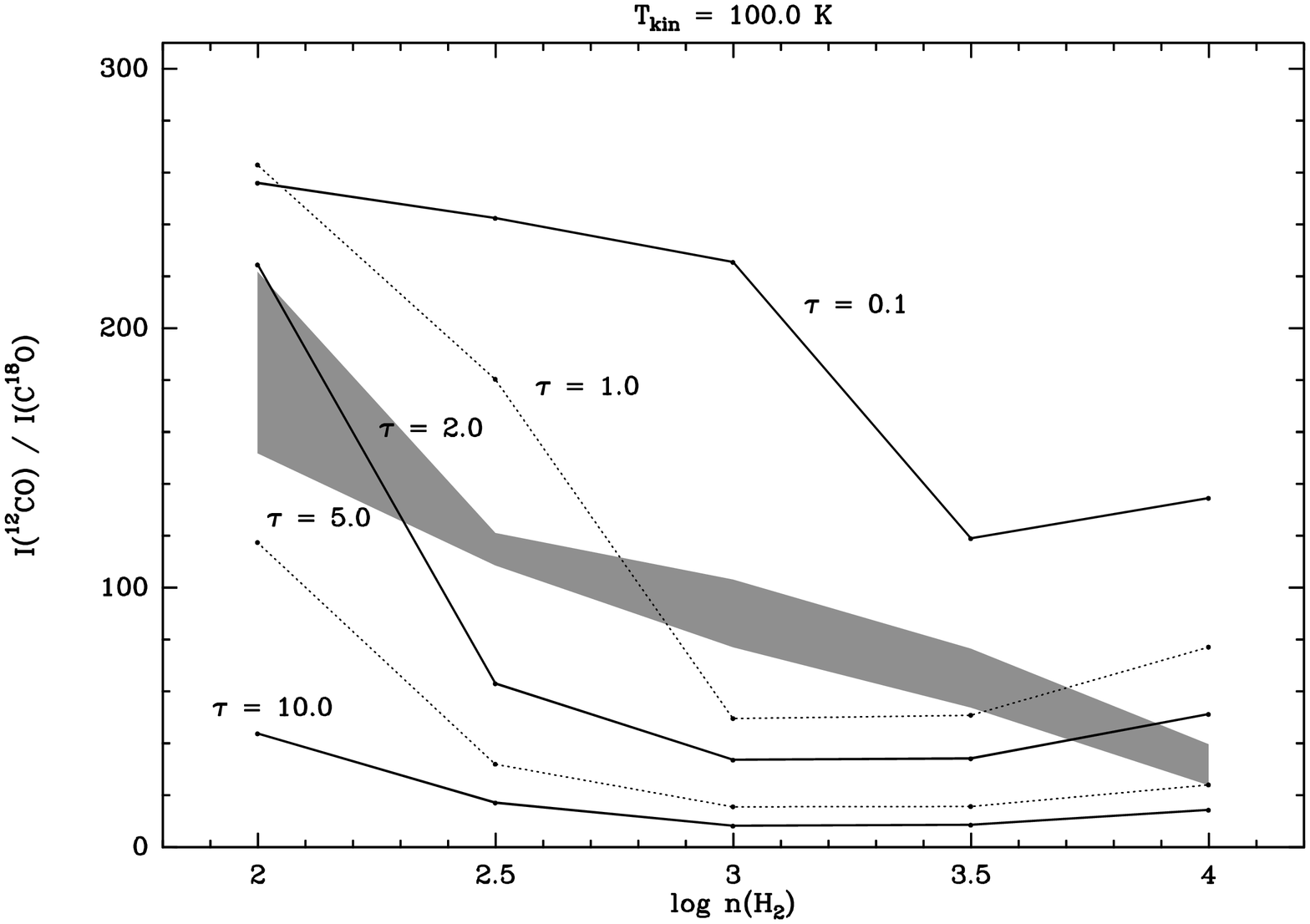,angle=-90,width=8.75cm}
\psfig{figure=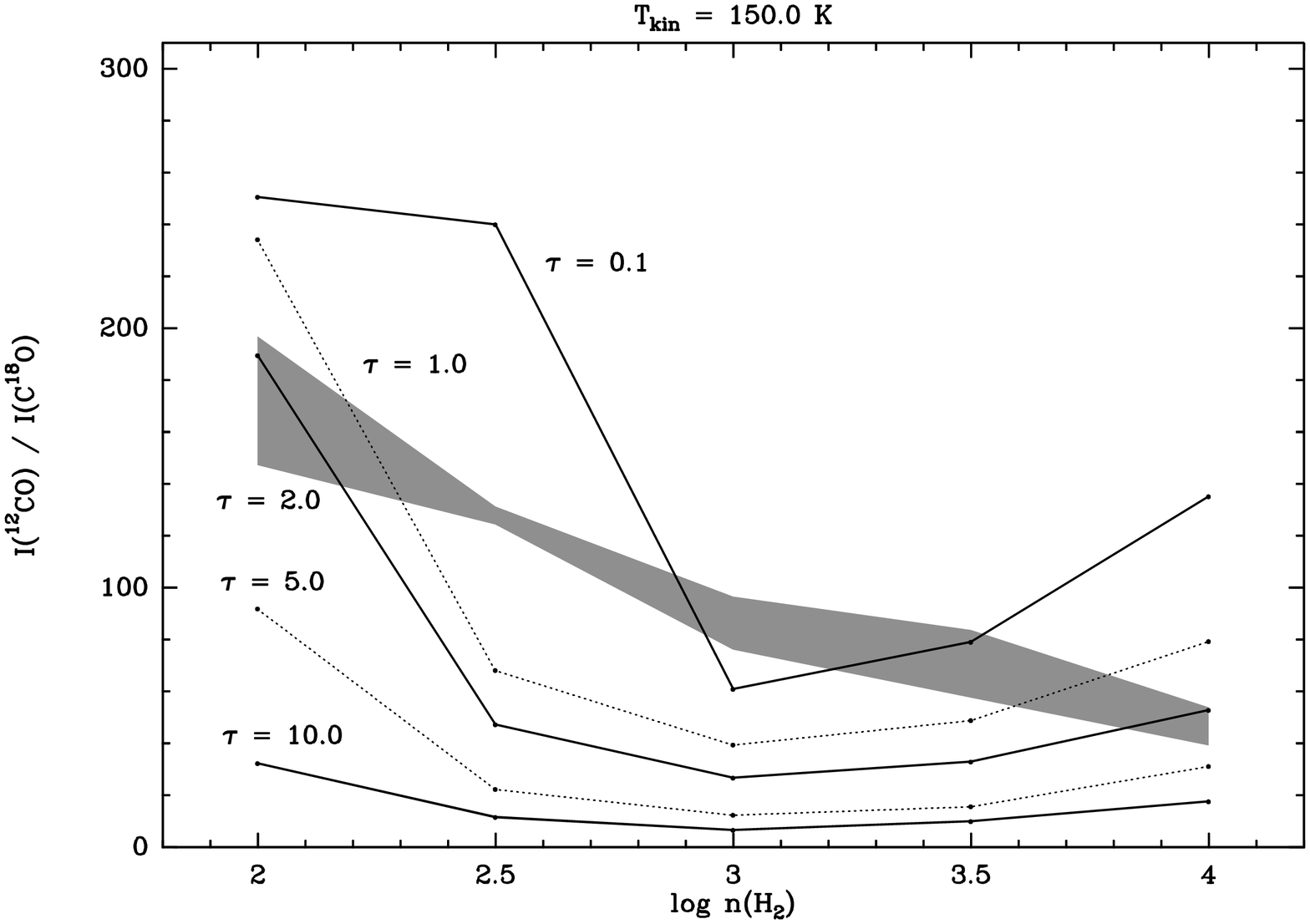,angle=-90,width=8.75cm}
\caption{The predicted $^{12}$CO/C$^{18}$O intensity ratios in the $J=1
\to 0$ transition as a function of the H$_2$ density for \Tkin{} =
50.0~K, 100.0~K, and 150.0~K from LVG calculations. The ratios are
determined for optical depths of 0.1, 1.0, 2.0, 5.0, and 10.0 in the
$^{12}$CO(1--0) transition, with [$^{12}$CO/H$_2$] = 10$^{-4}$ and
[$^{12}$CO/C$^{18}$O] = 250. The temperature of continuum background
radiation, $T_{\rm C}$, was chosen to be the cosmic background
radiation of 2.7~K.  The shaded area indicates the velocity gradient
range from 3~\kms\,pc$^{-1}$ to 6~\kms\,pc$^{-1}$, which is the most
likely value for the Galactic center region.} \label{plot_lvg_fits}
\end{figure}

For optically thin CO(1--0), the thermalization density is about
740~\pccm; the highest density for which CO intensities change
significantly with H$_2$ density is $\sim$\,10$^4$~\pccm. The fits were
always performed with the $^{12}$CO/H$_2$ ratio assumed to be
$10^{-4}$ and the C$^{18}$O/H$_2$ to be $4 \cdot 10^{-7}$, thus, the
$^{12}$CO/C$^{18}$O was assumed to be 250, since this is the commonly
accepted value for the Galactic center region \citep{WiMa92,WiRo94}.
See the caption of Fig.~\ref{plot_lvg_fits} for more details.

Because this investigation is focused on the optical depth, the program
was modified to fit an optical depth in the \mbox{$^{12}$CO(1--0)}
transition to given input parameters by varying the velocity gradient
and fixing all the other parameters ($n_{\rm H_2}$, \Tkin, $X$, and
$T_{\rm C}$).  The velocity gradient determined for the optical depths
of the \mbox{$^{12}$CO(1--0)} transition was taken as an input
parameter for the respective C$^{18}$O calculations. The calculated
optical depths for the \mbox{C$^{18}$O(1--0)} transition were always
$\ll 1$, confirming the notion that the \mbox{C$^{18}$O(1--0)}
transition is always optically thin. The results of the LVG
calculations are plotted in Fig.~\ref{plot_lvg_fits}.

In Section~\ref{COIntRat}, we distinguish four intensity ratio ranges:
low ratios of about 40 (a few positions), intermediate ratios of 60 to
80 (widespread), high ratios of 90 to 120 (several areas), and very
high ratios up to $>$\,200 (a few positions). From
Fig.~\ref{plot_lvg_fits}, one can see that high $^{12}$CO/C$^{18}$O
intensity ratios are usually correlated with lower optical depth in the
\mbox{$^{12}$CO(1--0)} transition. In addition, for higher kinetic
temperatures high $^{12}$CO/C$^{18}$O intensity ratios are more
restricted to low densities. However, these are only trends, and one
can find an appropriate optical depth to fit the required intensity
ratio for nearly every pair of H$_2$ density and \Tkin.  Therefore, one
needs an additional restriction on the possible range of  H$_2$ density
and \Tkin.  This can be the velocity gradient which, from the fits,
covers an unrealistically large range from 0.004~\kms\,pc$^{-1}$ to
462~\kms\,pc$^{-1}$. A reasonable gradient can be estimated from the
typical velocity extent and diameter of molecular clouds in the
Galactic center as it is found in the higher resolution maps of
$^{13}$CO(1--0) \citep{Bal+87} and \mbox{C$^{18}$O(1--0)}
\citep{Lin+95}.  With a typical velocity extent of 30~\kms{} to
50~\kms{} and a diameter of 2\arcmin{} to 5\arcmin, keeping in mind
that larger cloud diameters are correlated to larger velocity extents,
we estimate that the velocity gradient of Galactic center clouds
typically ranges from 3~\kms\,pc$^{-1}$ to 6~\kms\,pc$^{-1}$.  This
range is shown shaded in Fig.~\ref{plot_lvg_fits}.

The range of H$_2$ density and \Tkin{} is now rather strongly limited
and more definite parameters for the four ranges of intensity ratios
can be estimated. For all three \Tkin{} values, the allowed
$^{12}$CO/C$^{18}$O intensity ratio declines monotonically with
increasing H$_2$ densities. To get estimates for H$_2$ density and
\Tkin{} of the ratio ranges, we discuss each range separately taking
into account additional information from other observations:
\begin{enumerate}
\item For low intensity ratios of about 40, a fairly high H$_2$
      density is predicted. For \Tkin\,=\,50~K, $n_{\rm
      H_2}$\,$>$\,$10^{3.3}$~\pccm{} is required. For higher \Tkin,
      this value is even larger. \cite{Hut+93b} suggested the existence
      of two gas components: a cool (\Tkin\,$\approx$\,25~K) component
      with higher density and a hot (\Tkin\,$\approx$\,200~K) component
      with lower density. Because the positions studied by
      \citen{Hut+93b} were centered on the \mbox{CS(2--1)} peak
      positions of \cite{Bal+87}, they argued that CS in the cool and
      denser component should be thermalized, with H$_2$ densities
      above $1.8 \cdot 10^4$~\pccm.  Away from such peak positions,
      averaged over the 9\arcmin{}-beam, densities should be lower,
      perhaps by a factor of 5 to 10.  Similarly, \Tkin{} should be
      slightly higher. Therefore, we favor a beam averaged H$_2$
      density of $\sim$\,10$^{3.5}$~\pccm, a kinetic temperature of
      $\sim$\,50~K, and a \CO{12} optical depth of $\sim$\,3.0 for
      those regions showing low intensity ratios.
\item Intermediate ratios of 60 to 80 are most common, especially
      at $-$50~\kms{} and 0~\kms, and should, therefore, not have a
      density above the average. If the extent of the molecular
      emission along the line of sight is comparable to the
      longitudinal extent of the bulge region and if the scale height
      of the the molecular emission is $\sim$\,0\fdg{}2, the mean H$_2$
      density is (with a total gas mass of $3 \cdot 10^7$~\Msol,
      determined in Section~\ref{TotMolMass}) $\sim$\,150~\pccm.
      However, some clumping is expected. For \Tkin\,=\,50~K, the LVG
      results require $n_{\rm H_2}$\,$>$\,$10^{2.8}$~\pccm; for higher
      \Tkin, this value will be even larger. Therefore, we estimate
      that the most appropriate (and lowest possible) density for this
      intensity range is $n_{\rm H_2}$\,$\approx$\,$10^{3.0}$~\pccm, at
      \Tkin\,=\,50~K and $\tau_{\rm ^{12}CO}$\,$<$\,2.0.
\item For high ratios of 90 to 120, the LVG calculations require H$_2$
      densities of an order of magnitude lower than in case (1) for the
      same \Tkin{} values. Since \citep[following ]{Hut+93b} it is more
      likely that \Tkin{} is larger for gas with lower density, but
      since the ratios considered here are not the highest,
      we estimate that \Tkin{} is $\sim$\,100~K.  This implies an H$_2$
      density of $\sim$\,10$^{3.0}$~\pccm{} and $\tau_{\rm
      ^{12}CO}$\,$<$\,1.0. For Clump~2, this result is confirmed by
      \cite{StBa86} who found an optical depth of less than unity from a
      comparison of \mbox{$^{12}$CO(1--0)} to \mbox{$^{13}$CO(1--0)}
      emission using the usually assumed isotope ratio of
      $^{12}$C/$^{13}$C = 20 \citep{Pen80,WiMa92,WiRo94}.
\item For very high ratios up to over 200, the LVG results require an H$_2$
      density of $n_{\rm H_2}$\,$<$\,$10^{2.5}$~\pccm. We favor a
      \Tkin\,$\approx$\,150~K for this low density gas as the
      beam-averaged value of the hot (\Tkin\,$\approx$\,200~K) low
      density component found by \cite{Hut+93b}. This implies an H$_2$
      density of $\sim$\,10$^{2.0}$~\pccm{} and an optical depth of
      $\sim$\,2.0.  Such a high value of \Tkin{} is supported by
      calculations of \cite{Flo+95} who estimated
      \Tkin\,$\approx$\,500~K for the low density envelope of Sgr\,B2.
      Hence, even in our 9\arcmin{}-beam 150~K seems to be a reasonable
      value of \Tkin.
\end{enumerate}
We list our best estimates of the physical parameters in the four line
intensities ratio ranges in Table~\ref{TbPhysParEst}. Note that the large
beam {\it averages\/} over a wide range of environments. Thus, the derived
values indicate which conditions dominate within the 9\arcmin{}-beam. This
does not mean that they are exclusively present as can be seen in a
detailed analysis of high density peaks \citep{Hut+97}.

\begin{table}
\caption{Best estimates for beam averaged gas parameters for four
ranges of line intensity ratios, based on LVG calculations}
\label{TbPhysParEst}
\vspace{2mm}
\begin{tabular}{c|@{\hspace{1cm}}c@{\hspace{1cm}}c@{\hspace{1cm}}c}
$\boldmath I_{\bf ^{12}CO}/I_{\bf C^{18}O}$ & 
{\bf log$\boldmath (n_{\bf H_2})$} &
$\boldmath T_{\bf kin}$ &
$\boldmath \tau_{\bf ^{12}CO}$ \\
 & [cm$^{-3}$] & [K] & \\ 
\hline  & & & \\[-3mm] 
{\bf $\losim$ 40} &  3.5  & \enspace 50 & $\sim 3.0$ \\
{\bf 60 -- 80}    &  3.0  & \enspace 50 & $< 2.0$ \\
{\bf 90 -- 120}   &  3.0  &     100     & $< 1.0$ \\
{\bf up to 205}   &  2.0  &     150     & $\sim 2.0$ \\
\hline
\end{tabular}
\end{table}

If the molecular clouds are exposed to UV-radiation which can
dissociate the CO molecule, C$^{18}$O should be  more localized in the
cloud cores than $^{12}$CO because the self-shielding of the more
abundant $^{12}$CO is more effective \citep[see, e.g., ]{BaLa82}.
If so, the $^{12}$CO/C$^{18}$O ratios will be larger in the outer
layers of the clouds and, hence, the same intensity ratio corresponds
to a higher optical depth in $^{12}$CO. However, this effect should not
be large because the UV-radiation field in the Galactic center region,
away from Sgr\,B2, is not intense and a large part of the shielding
should be provided by dust. In addition, if this effect is present it
should affect high intensity ratios with weak undetected C$^{18}$O
emission the most. Thus, the true ratios in areas where UV dissociation
might be important are higher and our use of lower limits tends to
deemphasize this ``UV-radiation effect''.

So far, we have assumed that the $^{12}$CO and the C$^{18}$O emission
arises in the same volume. However, in a scenario of cool and dense
molecular cloud cores surrounded by a hotter and thinner molecular
component \cite{Hut+93b}, it is possible that the $^{12}$CO emission
originates in part or even completely in the thin component, while the
C$^{18}$O emission arises mostly from these clouds cores. This is the
case if the cloud cores are self-shielded in $^{12}$CO. Sgr\,B2 is, in
fact, known to have a hot low density envelope of molecular gas
\citep[see, e.g., ]{Hut+95}. Then, the line intensity of $^{12}$CO is
based on higher \Tkin{} and lower $n_{\rm H_2}$ than the line intensity
of C$^{18}$O. To model the resulting line intensity ratio of such a
scenario, we used the C$^{18}$O line intensities of our LVG calculations
determined for densities higher by 0.5 dex and for temperatures lower by
50~K compared to the values used for $^{12}$CO. It turns out that, for
the range of parameters considered above, the resulting ratios are
uniformly low (25\,--\,40). The dominating effect in this result is the
density gradient which causes a large reduction of the intensity ratio.
The temperature gradient causes a slight increase of the intensity ratio
which is an opposite but only marginal effect compared to the density
gradient. Obviously, such ratios are not compatible with our
observations. Therefore, our result that the optical depth of $^{12}$CO
is moderate or even low is confirmed and our assumption that the
$^{12}$CO and the C$^{18}$O emission arises in the same volume is
justified.

In summary, the LVG calculations confirm that the large scale
\mbox{$^{12}$CO(1--0)} emission in the Galactic center region is
dominated by emission with intermediate \mbox{($\tau = 1$\,--\,5)} or
low optical depths ($\tau < 1$). High optical depth emission ($\tau \ge
10$) commonly found in molecular clouds of the Galactic disk
\citep[e.g.\ ]{Phi+79} is restricted to very limited areas such as
Sgr\,B2.

\subsection{Estimates of the H$_{\it 2}$ Mass from CO Isotopomers} 

We now estimate the total H$_2$ mass of the inner Galactic center
region, i.e.\ the Galactic bulge. From the extent of the
\mbox{$^{12}$CO(1--0)} emission (see Fig.~\ref{12COContAllBit}) the
Galactic bulge is assumed to range from $l = -1$\fdg{}5 to +2\fdg{}25
and $b = -0$\fdg{}75 to +0\fdg{}75, corresponding to 555~pc $\times$
220~pc.

\subsubsection{Mass Estimate from the $^{12}$CO Standard Conversion Factor}

As a first step, the total H$_2$ mass is estimated from the
\mbox{$^{12}$CO(1--0)} luminosity determined from the data of
\cite{Bit87}. The SCF (Eq.~\ref{Eq-H2-from-CO}) gives the H$_2$ column
density. To obtain the total H$_2$ mass, we have to multiply this
result by the mass of a H$_2$ molecule and by the area each position
represents. This area is given by the spacing of the map, hence,
$0.\!\!^{\circ}125 \times 0.\!\!^{\circ}125$ which corresponds to
$3.274 \cdot 10^{39}$~cm$^2$ at 8.5~kpc. Therefore, we obtain:
\begin{eqnarray} \label{H2Mass-from-SCF}
{\cal M}_{\rm H_2}^{\rm ^{12}CO} & = & \frac{1.27 \cdot 10^3\,
      {\rm M_{\odot}}}{\rm K\ km\ s^{-1}} \cdot \nonumber \\ 
  & & \sum_{positions} \, \int 
      T_{\rm MB}^{\rm ^{12}CO(1-0)}(v) \, {\rm d}v
\end{eqnarray}
The total luminosity of the Galactic bulge in \mbox{$^{12}$CO(1--0)} is
$1.52 \cdot 10^5$~K\,\kms{} where the intensities are integrated over
the velocity range from $-$225.0 to +225.0~\kms{} and summed over the
area from $l = -1$\fdg{}5 to 2\fdg{}25 and from $b = -0$\fdg{}75 to
+0\fdg{}75. With this we obtain a mass of:
\begin{equation} \label{TotH2Mass12CO}
{\cal M}_{\rm H_2}^{\rm ^{12}CO}(^{l = -1.5\ {\rm to}\ +2.25}_{b = 
-0.75\ {\rm to}\ +0.75}) = 1.93 \cdot 10^8 \, {\rm M_{\odot}}
\end{equation}

\subsubsection{Mass Estimate from the C$^{18}$O Data} 

In the optically thin case, the column density of C$^{18}$O in the 
$J=0$ state can be determined from the integrated intensity as 
\citep[see Eq.~14.45 in ]{RoWi96}:
\begin{equation} \label{Eq-col-den-c18o-j0}
\bar{\cal N}_{J=0}^{\rm C^{18}O} = 
   \frac{1.21 \cdot 10^{14}\,{\rm cm}^{-2}}{\rm K\ km\,s^{-1}} \, 
   \int T_{\rm MB}^{\rm C^{18}O(1-0)} \, {\rm d}v 
\end{equation}
The ratio of $\bar{\cal N}_{J=0}^{\rm C^{18}O}$ to the total column
density of C$^{18}$O can be estimated from LVG calculations if we
assume reasonable average values for \Tkin{} and $\log (n_{\rm H_2})$.
Because \mbox{C$^{18}$O(1--0)} is optically thin, emission from
subthermally excited gas is too weak to be seen and $\log (n_{\rm
H_2})$ is $\grsim$\,3.0. About 60\% of the integrated intensity of
C$^{18}$O originates from positions in the $l,b,v$-space where the
$^{12}$CO/C$^{18}$O ratio is lower than 50. Compared with the
components identified in Section~\ref{LVGCalc}, it seems likely that
most of the C$^{18}$O emission originates in gas with $\log (n_{\rm
H_2})$\,$\approx$\,3.5 (at \Tkin\,$\approx$\,50~K and $\tau^{\rm
^{12}CO(1-0)}$\,$\approx$\,3.0).  This is in good agreement with
\cite[chap.\ 5]{Hut93}, who obtained $\log (n_{\rm
H_2})$\,$\grsim$\,3.7 for \CO{18} emission toward \mbox{CS(2--1)} peak
positions. For these values the LVG calculations with the appropriate
velocity gradients give $\bar{\cal N}^{\rm C^{18}O} \approx 6.5 \cdot
\bar{\cal N}_{J=0}^{\rm C^{18}O}$. This value varies by a factor of 2
if the density is modified to $\log (n_{\rm H_2})$\,=\,3.0 or 4.0 or if
\Tkin{} is increased. To obtain the total H$_2$ mass from $\bar{\cal
N}^{\rm C^{18}O}$, we take C$^{18}$O/H$_2 = 4 \cdot 10^{-7}$ (i.e.\
$^{12}$CO/H$_2 = 10^{-4}$), and then carry out the same steps as for
$^{12}$CO. Because the spacing of the map was 9\arcmin{} instead of
7\farcm{}5, the area each position represents is $0.\!\!^{\circ}15
\times 0.\!\!^{\circ}15 \hateq 4.715 \cdot 10^{39}$~cm$^2$. Therefore,
the total H$_2$ mass is given by:
\begin{eqnarray} 
{\cal M}_{\rm H_2}^{\rm C^{18}O} & = & \frac{1.56 \cdot 10^4 \,
      {\rm M_{\odot}}}{\rm K\ km\ s^{-1}} \cdot \nonumber \\  
  & & \sum_{positions} \, \int 
      T_{\rm MB}^{\rm C^{18}O(1-0)}(v) \, {\rm d}v
\end{eqnarray}
With the luminosity of \mbox{C$^{18}$O(1--0)} being 745~K\,\kms{} where
the intensities are integrated over the same velocity range as
$^{12}$CO and summed over the area from $l = -1$\fdg{}05 to 2\fdg{}25
and from $b = -0$\fdg{}75 to +0\fdg{}75, we find a total H$_2$ mass
of:
\begin{equation} \label{TotH2MassC18O}
{\cal M}_{\rm H_2}^{\rm C^{18}O}(^{l = -1.05\ {\rm to}\ +2.25}_{b = 
     -0.75\ {\rm to}\ +0.75}) = 1.16 \cdot 10^7 \, {\rm M_{\odot}}
\end{equation}
This value is more than an order of magnitude less than that determined
by the SCF from \mbox{$^{12}$CO(1--0)}. It is also lower than the
``weighted guess'' of about $0.8 \cdot 10^8 \, {\rm M_{\odot}}$
compiled by \cite{Gus89} from a collection of several observations (see
also Section~\ref{TotMolMass}).

However, we must take into account that there might be a considerable
contribution to the molecular mass from C$^{18}$O emission which is
below our detection limit (see Section~\ref{COIntRat}). To determine an
upper limit, we determine the luminosity of \mbox{C$^{18}$O(1--0)} with
a 1$\sigma$-threshold which gives us the absolute upper limit allowed
by the data. Certainly, the luminosity of \mbox{C$^{18}$O(1--0)} is
expected to be lower because the intensity of C$^{18}$O is not
everywhere as high as 1$\sigma$ where it is not detected. To avoid
areas with {\it no\/} molecular gas, we use the presence of $^{12}$CO
emission as a constraint for the integration range in C$^{18}$O.  As
mentioned in Section~\ref{12COLVPlots} of \pageref{12COLVPlots}, the
$^{12}$CO emission has a constant velocity width of about 375~\kms, the
central velocity of which can be described by the relation with $v_{\rm
center}(l) = l \cdot 41.4$~\kms{}/arcdeg + 7.2~\kms{} for $l \le
2$\fdg{}0 and $v_{\rm center}(l) = 90.0$~\kms{} for $l > 2$\fdg{}0.
With this, the upper limit of the \mbox{C$^{18}$O(1--0)} luminosity is
4\,100~K\,\kms.  Then, the upper limit of the total H$_2$ mass from
\mbox{C$^{18}$O(1--0)} emission allowed by the data is:
\begin{equation} 
{\cal M}_{\rm H_2}^{\rm C^{18}O}(^{l = -1.05\ {\rm to}\ +2.25}_{b = 
     -0.75\ {\rm to}\ +0.75}) < 6.41 \cdot 10^7 \, {\rm M_{\odot}}
\end{equation}
Even this limit is a factor of 3 lower than the total H$_2$ mass
determined with the SCF and still slightly lower than the ``weighted
guess'' by \cite{Gus89}.

A larger $^{12}$CO/C$^{18}$O ratio due to selective UV dissociation of
C$^{18}$O would increase ${\cal M}_{\rm H_2}^{\rm C^{18}O}$ linearly;
conversely a $^{12}$CO/H$_2$ ratio $>$\,10$^{-4}$ due to higher
metallicity in the Galactic center would decrease {\it all\/} masses
derived from CO isotopomers. For example, \cite{Irv+85} find $5 \cdot
10^{-4}$ in the envelope of the evolved carbon star IRC+10216 and
\cite{Ari+96} derive metallicity-dependent CO-to-H$_2$ conversion
factors. Both effects might add an uncertainty of a factor of $\sim$\,2
to the ${\cal M}_{\rm H_2}^{\rm C^{18}O}$ mass, but will cause changes
in opposite directions. Considering that the scaling of $\bar{\cal
N}_{J=0}^{\rm C^{18}O}$ to $\bar{\cal N}^{\rm C^{18}O}$ is also
uncertain by about a factor of 2, we estimate that the ${\cal M}_{\rm
H_2}^{\rm C^{18}O}$ is uncertain by a factor of $\sim$\,3. Thus, it is
formally possible (but unlikely) to raise the upper limit of ${\cal
M}_{\rm H_2}^{\rm C^{18}O}$ allowed by the data sufficiently to be
consistent with the mass obtained from $^{12}$CO by the SCF. However,
because this formal upper limit is for sure significantly larger than
the true value, any consistency with the SCF can be ruled out.

\subsubsection{The Thin Gas Contribution} 

We find a molecular gas component with low H$_2$ densities and high
kinetic temperatures to be widespread in agreement with \cite{Oka+97}. 
In this component, the CO level populations are likely to be
subthermal, as indicated by the high $^{12}$CO/C$^{18}$O ratio and the
LVG calculations. In this case, emission of C$^{18}$O is very difficult
to detect since the subthermal excitation is correlated with low gas
densities and low column densities. Indeed about 60\% of the C$^{18}$O
emission originates in $l,b,v$-channels where the $^{12}$CO/C$^{18}$O
ratio is below 50 whereas only 10\% of it originate in $l,b,v$-channels
where the $^{12}$CO/C$^{18}$O ratio is above 90. The non-detection of
C$^{18}$O places a limit on the possible H$_2$ column density (and
mass) in this region (see above). Using the physical conditions given
by the LVG-model for high $^{12}$CO/C$^{18}$O ratios, we can achieve a
better estimate of the (possibly significant) contribution of this
``thin gas component'' missing from the C$^{18}$O mass budget.

We assume that the $^{12}$CO emission which appears at positions in the
$l,b,v$-space where no C$^{18}$O emission is detected originates in the
thin gas component. This criterium is reasonable but not without
problems, since (1) some thermalized C$^{18}$O emission might be
present below our detection limit and (2) even if we detect C$^{18}$O,
it is not guaranteed that {\it all\/} the $^{12}$CO emission originates
in thermalized molecular gas from which the C$^{18}$O emission arises.
However, our LVG calculations predict subthermal excitation conditions in
regions of high $^{12}$CO/C$^{18}$O ratios and moderate optical depths.
As discussed above, this excludes that $^{12}$CO emission from hot, low
density gas can shield the dense cores completely.

\begin{table*}
\caption{The gas mass in the central 600~pc of the Galactic center from
different tracers} \label{GasMassSum}
\begin{tabular}{lrl}
{\bf Tracer} & \multicolumn{1}{c}{$\boldmath {\cal M}_{\bf mol}$}
 & {\bf Reference} \\[1mm] 
\hline & & \\[-3mm] 
$^{12}$CO(1--0) (SCF)      &    $2.8 \cdot 10^8 \, {\rm M_{\odot}}$
 & This work, Eq.~\ref{TotH2Mass12CO}$^{\rm a}$ \\ 
C$^{18}$O(1--0)            &          $1.7 \cdot 10^7 \, {\rm M_{\odot}}$
 & This work, Eq.~\ref{TotH2MassC18O}$^{\rm a}$ \\ 
$^{12}$CO(1--0) (thin gas) &     $\sim 1.0 \cdot 10^7 \, {\rm M_{\odot}}$
 & This work, Eq.~\ref{H2Mass-Thin12CO}$^{\rm a}$ \\ 
Dust\,/\,IRAS              &          $3.6 \cdot 10^7 \, {\rm M_{\odot}}$
 & \cite{CoLa89} \\ 
Dust 800\,$\mu$m           &        $> 0.4 \cdot 10^7 \, {\rm M_{\odot}}$
 & \cite{LiCa94} \\ 
Dust\,/\,COBE              & 3.1\,--\,$7.0 \cdot 10^7 \, {\rm M_{\odot}}$
 & \cite{Sod+94}\,/\,data of \cite{Bit87}$^{\rm a}$ \\ 
0.1--1.0 GeV $\gamma$-rays &        $< 5.8 \cdot 10^7 \, {\rm M_{\odot}}$
 & \cite{Bli+85} \\ 
\hline & & \\[-3mm] 
\multicolumn{3}{l}{\footnotesize $^{\rm a}$ Divided by 0.7 to take into 
account that only 70\% of the total mass is found in hydrogen.} \\ 
\end{tabular}
\end{table*}

In estimating the mass of the thin gas component, we considered the
complete range of emission from $-$225~\kms{} to +225~\kms{} for
$^{12}$CO and restricted the integration to positions in the
$l,b,v$-space where the C$^{18}$O emission is below $2\,\sigma_{\rm
C^{18}O}$, i.e.\ below 0.072~K. This procedure required the $^{12}$CO
data to be sampled on the same grid in the $l,b,v$-space as the
C$^{18}$O data. Thus, the data cube with the $^{12}$CO data of
\cite{Bit87} resampled to the slightly lower velocity and spatial
resolution of the C$^{18}$O spectra (see Section~\ref{COIntRat}) was
used. From this, the $^{12}$CO luminosity of areas where the C$^{18}$O
emission is too weak to be detected is $7.78 \cdot 10^4$~ K\,\kms.
This is roughly $1/2$ of the total luminosity of $^{12}$CO. Because we
expect the optical depth of $^{12}$CO to be moderate ($\tau_0 \losim 1$) 
but not $\ll 1$, we use the approximation 
\begin{equation} \label{Eq-Tex-tau-Tmb} 
f_{\rm beam} \, T_{\rm ex} \int \tau(v) \, {\rm d}v 
      \cong \frac{\tau_0}{1 - e^{ - \tau_0}} 
      \int T_{\rm MB}(v) \, {\rm d}v
 \end{equation}
which is accurate to 15\% for $\tau_0 < 2$, and it always 
overestimates $\bar{\cal N}$ when $\tau_0 > 1$ \citep{RoWi96}. Then, we
obtain for the column density of $^{12}$CO in the state $J=0$:
\begin{eqnarray}
\bar{\cal N}_{J=0}^{\rm ^{12}CO_{thin}} & = & \frac{1.15 \cdot 
       10^{14}\,{\rm cm}^{-2}}{\rm K\ km\,s^{-1}} \cdot \nonumber \\  
 &   & \frac{\tau_0^{\rm ^{12}CO(1-0)}}{1 - 
       e^{ - \tau_0^{\rm ^{12}CO(1-0)}}} \, 
       \int T_{\rm MB}^{\rm ^{12}CO(1-0)} \, {\rm d}v 
\end{eqnarray}
Assuming \Tkin\,$\approx$\,150.0~K and $n_{\rm
H_2}$\,$\approx$\,$10^{2.5}$~\pccm{} as a good approximation to the hot
low density component, the LVG calculations (Section~\ref{LVGCalc})
yield $\bar{\cal N}^{\rm ^{12}CO_{thin}} \approx 5.75 \cdot \bar{\cal
N}_{J=0}^{\rm ^{12}CO_{thin}}$ and $\tau_0^{12}/\left( 1 - \exp ( -
\tau_0^{12} ) \right) \approx 1.78$.  Assuming again a $^{12}$CO/H$_2$
ratio of $10^{-4}$, the H$_2$ mass is:
\begin{eqnarray} 
{\cal M}_{\rm H_2}^{\rm ^{12}CO_{thin}} & = & \frac{93.8 \, 
       {\rm M_{\odot}}} {\rm K\ km\ s^{-1}} \cdot \nonumber \\ 
 &   & \sum_{positions} \, \int 
       T_{\rm MB}^{\rm ^{12}CO(1-0)}(v) \, {\rm d}v
\end{eqnarray}
Comparing this equation with Eq.~\ref{H2Mass-from-SCF}, it is obvious
that in case of low or intermediate optical depth the same
\mbox{$^{12}$CO(1--0)} luminosity corresponds to an H$_2$ mass which is
more than an order of magnitude lower.  With the above estimate of the
\mbox{$^{12}$CO(1--0)} luminosity of the ``thin gas component'' we
determine:
\begin{equation} \label{H2Mass-Thin12CO}
{\cal M}_{\rm H_2}^{\rm ^{12}CO_{thin}}(^{l = -1.05\ {\rm to}\ +2.25}_{b = 
     -0.75\ {\rm to}\ +0.75}) = 0.73 \cdot 10^7 \, {\rm M_{\odot}}
\end{equation} 
This is $\sim$\,60\% of the mass determined from C$^{18}$O. 

This mass will be larger if the portion of $^{12}$CO emission which
originates in thin gas is higher, or if the $^{12}$CO/H$_2$ ratio in
this component is lower than $10^{-4}$, because of less effective
shielding in low-density gas, leading to enhanced CO dissociation.
Thus, we estimate that the molecular mass in the thin gas component
might be as high as twice the value calculated.  Therefore, this
component may contain as much gas as the denser and cooler cloud cores
traced by C$^{18}$O.

\section{Discussion} \label{Discuss}

\subsection{The Molecular Mass in the Galactic Bulge} \label{TotMolMass} 

\subsubsection{Mass Estimates from other Tracers} 

To better assess the uncertainties of our total molecular mass
estimate, we compare the values obtained from CO data with results
obtained from other tracers.
 
\cite{CoLa89} analyzed IRAS data of thermal dust emission in the inner
500~pc of the Galactic Center. Using \Tdust\,=\,27~K and a metallicity
$Z$\,=\,2\,$Z_{\odot}$, they found a total gas mass of $3.6 \cdot 10^7
\, {\rm M}_{\odot}$.

\cite{LiCa94} presented an 800\,$\mu$m survey of the inner 1\fdg{}5
$\times$ 0\fdg{}2 which is sensitive to cooler dust than IRAS.
However, due to a small chopper amplitude, their estimate is limited to
small scale structure.  They determined a lower limit to the total gas
mass of $0.4 \cdot 10^7 \, {\rm M}_{\odot}$. One half of this mass
arises from the Sgr\,B2 region. From the mass determined from
C$^{18}$O, we estimate that \citen{LiCa94} recorded about 25\% of the
total gas in the region mapped and about 50\% toward Sgr\,B2. When
compared to our data, we find that the distribution of the dust is
better correlated with the C$^{18}$O emission than with the $^{12}$CO
emission, especially near Sgr\,B2.

In an analysis of Galactic plane surveys from the Diffuse Infrared
Background Experiment (DIRBE) of the Cosmic Background Explorer (COBE)
at 140\,$\mu$m and 240\,$\mu$m, \cite{Sod+94} obtained dust-to-gas
ratios ($d/g$) along the Galactic plane. If one assumes a unique
$I_{\rm CO}/{\cal N}_{\rm H_2}$ conversion factor and compares dust
emission to $^{12}$CO and H\,I data, one finds that the ratio in the
inner 4\degr{}, i.e.\ the Galactic bulge, decreases by a factor of
$2-3$ relative to the inner disk.  The dust-to-gas ratio also
depends on metallicity \citep{Mez+86} which is $2-3$ times higher in
the Galactic bulge than in the inner disk \citep{CoLa89}. Then, the
dust-to-gas ratio in the Galactic bulge would actually be lower by a
factor $4-9$ than in the disk.  \cite{Sod+94} proposed that the ${\cal
N}_{\rm H_2}/I_{\rm CO}$ ratio is also lower by a factor of $4-9$.
Using this conversion factor, we obtain a total H$_2$ mass of 2.2 to
$4.9 \cdot 10^7 \, {\rm M}_{\odot}$ from the \mbox{$^{12}$CO(1--0)}
data of \cite{Bit87}, in good agreement with our C$^{18}$O result.

An independent mass estimate based on $\gamma$-rays from 0.1~GeV to
1.0~GeV that are produced when cosmic rays interact with the
interstellar gas \citep{Bli+85} gives an upper limit to the gas mass of
the Galactic bulge of $5.8 \cdot 10^7 \, {\rm M}_{\odot}$. However,
\cite{SiBl87} found that the $\gamma$-ray flux in the range from 1~GeV
to 5~GeV is compatible with the $^{12}$CO(1--0) luminosity and the SCF
if the cosmic-ray density ($> 1$~GeV) is the same as that found in the
Galactic plane by \cite{Blo+86}.

\subsubsection{A Weighted Best Estimate}

In Table~\ref{GasMassSum}, we summarize the values for the total
molecular mass in the Galactic bulge from different tracers. The
estimates of the total mass in the molecular phase based on dust and on
C$^{18}$O agree reasonably well, especially if one takes the thin gas
component into account. The $^{12}$CO data yield a much higher mass if
the SCF is used.

The previous weighted guess was $8 \cdot 10^7 \, {\rm M_{\odot}}$
\citep{Gus89}, including the mass determined from $^{12}$CO with the
SCF. There is now additional evidence for an even lower mass, from our
C$^{18}$O data and the COBE data presented by \cite{Sod+94}. In
addition, we have shown that the SCF {\it cannot\/} be applied to the
Galactic center region (see also next section). Therefore, our {\em
weighted best estimate\/} for the molecular mass in the Galactic bulge
is:
\begin{equation}
{\cal M}_{\rm mol} = (3^{+2}_{-1}) \cdot 10^7 \, {\rm M_{\odot}}
\end{equation}
The errors include the results of mass determinations from C$^{18}$O
data, $\gamma$-ray flux, and dust emission, but exclude the $^{12}$CO
SCF value.  They are also compatible with an additional hot low density
component of similar mass as the gas traced by C$^{18}$O.

\subsection{The $^{\it 12}$CO Conversion Factor toward the Galactic Bulge} 
\label{12COConvFac}

We have shown that the SCF gives a column density of molecular
hydrogen, $\bar{\cal N}_{\rm H_2}$, that is too high by an order of
magnitude in the Galactic bulge. This may have far-reaching
consequences for the interpretation of galactic and extragalactic data
and, hence, warrants a more detailed investigation.

The SCF, obtained by \cite{Str+88}, is based on two major assumptions:
a) the emission of \mbox{$^{12}$CO(1--0)} is optically thick and b) the
molecular clouds are virialized. A comparison of
Eq.~\ref{Eq-tcol-12co-thin} for $\bar{\cal N}_{\rm ^{12}CO}$ in case of
intermediate and low optical depths with Eq.~\ref{Eq-H2-from-CO}
(assuming \CO{12}/H$_2$ = 10$^{-4}$) for $\bar{\cal N}_{\rm ^{12}CO}$
in the optically thick case (SCF) shows that if one erroneously assumes
optically thick $^{12}$CO emission and applies the SCF
(Eq.~\ref{Eq-H2-from-CO}), one overestimates the total H$_2$ mass,
${\cal M}_{\rm H_2}$, by a factor of up to 80.  This illustrates that
only a few percent of $^{12}$CO molecules emitting under optically thin
conditions can contribute a large fraction of the $^{12}$CO luminosity
and, hence, strongly influence the estimates of gas masses.  Therefore,
the areas of low or intermediate optical depth, that dominate the
$^{12}$CO emission in the Galactic center over large scales, are at
least partly responsible for the overestimation of the molecular mass
by the SCF.

One reason for the anomalously low optical depths of the $^{12}$CO
emission are the large linewidths of Galactic center clouds which are
an order of magnitude larger than those found for disk Giant Molecular
Clouds (GMCs). Thus, the optical depths per column density are an order
of magnitude lower in the Galactic Center than in Galactic disk
clouds.

Furthermore, the assumption that the molecular clouds are virialized
may be not fulfilled in many cases.  Molecular clouds can only be
virialized if their mean density, $n_{\rm H_2}$, exceeds $10^4\ {\rm
cm}^3\ \left( 75\,{\rm pc} \, / \, {\rm D_{GC}} \right)^{1.8}$ where
\dgc{} is the distance from the Galactic center \citep[see, e.g.,
]{GuDo80}.  Therefore, the densities must be 2\,000 -- 5\,000~\pccm{}
in the outer bulge (D$_{\rm GC} > 100$~pc) and 10\,000~\pccm{} and more
in the inner bulge (D$_{\rm GC} < 100$~pc) to be stable. Since \CO{18}
is collisionally excited and, thus, detectable at these densities, we
expect that the molecular gas in virialized clouds is traced completely
by C$^{18}$O emission. With a mean density of the gas emitting
C$^{18}$O of about 3\,000~\pccm{}, the dense cores (which are the
strongest sources in C$^{18}$O) should indeed be virialized. This is
supported by the fact that mass estimates based on the SCF are closer
to the \CO{18} masses toward molecular peaks:  {\em Globally\/} the
discrepancy is about a factor of 16, dropping to 5--6 for Sgr\,B2 and
Sgr\,A, and becoming even smaller when the peaks are investigated with
higher angular resolution \citep[see, e.g., ]{Mau+89}.

For non-virialized gas, the $^{12}$CO line width traces not the
molecular mass of the cloud but the dynamical mass which causes the
gravitational potential the molecular cloud is moving through.
\cite{Dow+93} find that the standard conversion factor overestimates
the true gas mass by a factor of $\sqrt{{\cal M}_{\rm stars}/{\cal
M}_{\rm H_2}}$ if the average stellar mass density exceeds the gas mass
density. With $<$1\% of the mass of the Galactic bulge in the gas phase
\citep[see, e.g., ]{Gus89}, the mass determined from the $^{12}$CO
emission with the SCF should be too high by a factor 10, i.e.\ the true
value should be $2.8 \cdot 10^{7} \, {\rm M_{\odot}}$, in agreement
with the weighted best estimate derived in Section~\ref{TotMolMass} if
all the molecular gas is located in non-virialized clouds.
However, such a scenario implies that all the molecular gas fulfills
the other precondition of the SCF, namely that the $^{12}$CO emission is
optically thick. From our LVG calculations, this is not the case.
Thus, a significant part of the dense cores, containing a major
fraction of the total mass, must be virialized even though the
virialization density is rather high, in particular higher than the
thermalization density. Therefore, the non-virialization of
molecular clouds is an important but not the only reason for the
overestimate of the molecular mass from the $^{12}$CO luminosity by the
SCF. On the other hand, the fact that, nearly everywhere in the Galactic
center region, the virialization density is higher than the
thermalization density implies that the unbound gas cannot simply be
identified with the ``thin gas'' component discussed above.

In summary, the $^{12}$CO luminosity is larger than predicted by the
SCF since:
\begin{enumerate}
\item The $^{12}$CO emission on large scales is dominated by lines that
      are not optically thick.
\item A considerable fraction of the gas is not virialized.
\end{enumerate}
Both effects tend to affect the same gas component, and we stress that
the main cause of the difference in mass estimates from the two CO
isotopomers is not the emission from the peaks but the extended
emission from $^{12}$CO of rather low optical depth, in particular the
hot low density component (thin gas). Therefore, it is {\it not\/}
correct just to apply a modified conversion factor to the Galactic
center region as done by \cite{Ari+96} and \cite{Oka+97}. The point is
that in the Galactic center region the $^{12}$CO luminosity has
{\it no\/} direct causal connection with the molecular mass. This is
shown by the different line shapes of \mbox{$^{12}$CO(1--0)} and
\mbox{C$^{18}$O(1--0)} (Fig.~\ref{VerglSpectra}) and by the large
variation of the $^{12}$CO/C$^{18}$O ratio (Fig.~\ref{RatioContChan}).

\subsection{Consequences for other Galactic Nuclei} \label{ComExtGal} 

In our Galaxy, the Galactic center is a unique object displaying many
features not present in the Galactic disk. Thus, large scale physics and
properties found in detailed investigations of our Galactic center
region should be compared to central regions of other galaxies,
especially of spirals similar to ours.

\subsubsection{A Comparison with IC\,342}

The late-type spiral galaxy IC\,342 is thought to be similar to the
Milky Way.  For our comparison, we use the \mbox{$^{12}$CO(1--0)}
observations of the central region in IC\,342 obtained by \cite{Ish+90}
with the Nobeyama Millimeter Array (NMA). The map has an angular
resolution of 2\farcs{}4, corresponding to 22.5~pc at a distance to
IC\,342 of 1.85~Mpc \citep{MCa89}, which is identical to the linear
resolution of our C$^{18}$O and $^{12}$CO data. The extent of its
central region is 62\farcs{}5 $\times$ 37\farcs{}1 (560~pc $\times$
335~pc), very close to the size of our Galactic bulge.

From Fig.~2(a) of \cite{Ish+90}, the molecular gas traces two narrow
ridges which form a small-scale spiral in the bulge and are associated
with shock waves. Molecular condensations of similar structure and
extent as Sgr\,A, Sgr\,B2, Sgr\,C or the 1.5-complex (Sgr\,D region)
can be recognized. From the SCF, \citen{Ish+90} determine the molecular
mass of the central region of IC\,342 as $2.6 \cdot 10^8 \, {\rm
M}_{\odot}$. If, however, the mass in IC\,342 determined by this
factor is as greatly overestimated as in the Galactic center region,
the true molecular mass of IC\,342 should be $\sim 2.8 \cdot 10^7 \,
{\rm M}_{\odot}$, nearly the same as the weighted best estimate for our
Galactic bulge.

In a detailed analysis of the central region of IC\,342 based on the
high density tracer \mbox{HCN(1--0)} and \mbox{$^{12}$CO}
\citep{Ish+90}, \mbox{$^{13}$CO} \citep{Ish+91,TuHu92} and NH$_3$
\citep{Ho+90} data, \cite{Dow+92} concluded that the $^{12}$CO and
$^{13}$CO emission does not trace the same material, postulating at
least three gas components: low density gas at $150 - 300$~\pccm{}
accounts for most of the \CO{12}, moderate density gas at $500 -
3\,000$~\pccm{} for the \CO{13}\ and high density cores $\grsim
10^4$~\pccm{} for the HCN. From NH$_3$, the  kinetic temperature for
moderate and high densities is $50 - 70$~K. This picture closely
resembles our results for the Galactic center region (see this paper
for the low and moderate density component; high density cores traced
by the CS survey of \cite{Bal+87} are investigated in detail by
\cite[chapt.\ 5]{Hut93}; NH$_3$ observations are described in
\cite{Hut+93b}).

However, \cite{Dow+92} stated, without reference to method, that the
low density gas should have rather low kinetic temperatures of
$\sim$\,30~K.  But low density gas cools less effectively than denser
material that, in addition to cooling by line radiation, can transfer
energy to dust grains more easily because it is better coupled to dust
grains by collisions.  Therefore, the thin gas which should be heated
strongly by shocks is more likely to have higher kinetic temperatures
than the denser gas component, as it is found for our Galactic center
region.

\subsubsection{The Starburst Galaxy NGC\,253}

In contrast to the rather quiescent state of our Galaxy and of IC\,342,
the nearly edge-on spiral galaxy NGC\,253 \citep[$i = 78$\degr,
]{Pen81} is a good example for a nuclear starburst \citep{Rie+88}. The
molecular distribution and mass of this nearby ($D = 2.5$~Mpc) galaxy
was investigated in detail by \cite{Mau+96}. They found that the
molecular gas mass in the central condensations of NGC\,253 is also
overestimated from an application of the SCF.  They determined a mass
of about $5 \cdot 10^7 \, {\rm M}_{\odot}$ from C$^{18}$O lines and
mm-wave dust continuum for the central 300~pc whereas the SCF yields a
result which is higher by a factor of $\sim$\,6. They conclude that the
SCF cannot be applied to the central region of NGC\,253 for the same
reasons we found for our Galactic center region.

This mass determination is confirmed by \cite{Pag+95} who investigated
dense clouds in the central region of NGC\,253 in the \mbox{HCN(1--0)}
line using the Nobeyama Millimeter Array (NMA). From molecular
excitation models, they found a total mass of molecular gas within the
central 400~pc of $1.4 \cdot 10^8 \, {\rm M}_{\odot}$, in good
agreement with \cite{Mau+96} who determined $1.3 \cdot 10^8 \, {\rm
M}_{\odot}$.

\subsubsection{Other Galaxies}

For IC\,694 and the mergers NGC\,1614 and Arp\,220, \cite{Shi+94}
argued that the SCF cannot be applied to the central regions. They
determined the total mass present in the central regions and found in
all three galaxies that the H$_2$ masses derived from the SCF are
higher than the total mass including stars which is obviously
impossible. Together with the results previously mentioned and our
result for the central region of our Galaxy, this shows that the
non-applicability of the SCF is probably a general property of the
central region of galaxies.

The existence of a hot low density gas component in the central region
was also favored by \cite{Aal+95} for a sample of 32 external galaxies
consisting of starburst galaxies, interacting galaxies and two
quiescent systems. From their single-dish observations \citen{Aal+95}
concluded that most of the $^{12}$CO emission originates in this low
density component, under conditions of only moderate optical depth, and
found $^{12}$CO/C$^{18}$O ratios that are significantly above the
values in the disk of the Milky Way, although not as high as the most
extreme ratios found for our Galactic Center region. However, this is
readily explained by the fact that their beam averages over the dense
and the thin component, while the different components are at least
partly resolved toward the Galactic center region. This indicates that
the existence of a hot low density gas component which dominates the
$^{12}$CO emission is also very likely to be a general property of the
central region of galaxies.

\section{Conclusions} \label{Conclu} 

From the large scale C$^{18}$O(1--0) Galactic Center Survey, presented
in Paper~I, we obtain the following results:
\begin{enumerate}
\item In the Galactic center region, the standard conversion factor
      (SCF) of \cite{Str+88} which relates the H$_2$ column density to
      the integrated intensity of the \mbox{$^{12}$CO(1--0)} line,
      $I_{\rm ^{12}CO}$, is {\it not\/} valid. The SCF overestimates
      $\bar{\cal N}_{\rm H_2}$ by a factor of $\sim 9$.  This
      overestimate is caused by the high luminosity of \CO{12} lines of
      moderate and low optical depth from a rather small but widespread
      fraction of the gas and the non-virialization of a considerable
      fraction of the molecular gas in the gravitational potential of
      the Galactic center region. Therefore, also one can not apply a
      modified conversion factor to the Galactic center region since
      the $^{12}$CO luminosity is not connected in a direct way with
      molecular mass. In particular, this is shown by the different
      line shapes of \mbox{$^{12}$CO(1--0)} and \mbox{C$^{18}$O(1--0)}
      and by the large variation of the $^{12}$CO/C$^{18}$O integrated
      intensity ratio.
\item The integrated intensity ratio of $^{12}$CO/C$^{18}$O in the
      Galactic Center region is generally higher than the value of
      about 15 which is expected from the ``standard'' conversion
      factors. In the 9\arcmin{}-beam, this ratio is at least of the
      order 40, mostly of the order of 60 to 80, in several
      $l,b,v$-areas of the order 90 to 120, and toward a few positions
      up to $>$\,200.
\item From LVG calculations, we find that the large scale $^{12}$CO 
      emission in the Galactic Center region is dominated by
      emission with intermediate \mbox{($\tau = 1$\,--\,5)} or low
      optical depths ($\tau < 1$). High optical depth emission ($\tau
      \ge 10$) is restricted to very limited areas such as Sgr\,B2.
      Specifically, we find the following conditions:
      \begin{enumerate}
      \item For $^{12}$CO/C$^{18}$O integrated intensity ratios, 
            $\losim 40$: \\
            \makebox[30mm][l]{$n_{\rm H_2} \sim 10^{3.5}$~\pccm,}
            \makebox[22mm][l]{$T_{\rm kin} \sim 50$~K,}
            \makebox[13mm][l]{$\tau \sim 3.0$}
      \item For the most common ratios, 60 to 80: \\
            \makebox[30mm][l]{$n_{\rm H_2} \sim 10^{3.0}$~\pccm,}
            \makebox[22mm][l]{$T_{\rm kin} \sim 50$~K,}
            \makebox[13mm][l]{$\tau < 2.0$}
      \item For high ratios, 90 to 120, found, e.g., in the Sgr\,D region 
            and Clump~2: \\
            \makebox[30mm][l]{$n_{\rm H_2} \sim 10^{3.0}$~\pccm,}
            \makebox[22mm][l]{$T_{\rm kin} \sim 100$~K,}
            \makebox[13mm][l]{$\tau < 1.0$}
      \item For very high ratios up to 200: \\
            \makebox[30mm][l]{$n_{\rm H_2} \sim 10^{2.0}$~\pccm,}
            \makebox[22mm][l]{$T_{\rm kin} \sim 150$~K,}
            \makebox[13mm][l]{$\tau \sim 2.0$}
      \end{enumerate}
\item We estimate the total molecular mass of the Galactic bulge based
      on our C$^{18}$O data, on the $^{12}$CO data of \cite{Bit87}, and
      on dust measurements. A molecular gas component of low density
      (thin gas component) contributes a probably considerable amount
      of molecular mass. The {\em weighted best estimate\/} for the
      total molecular mass is ${\cal M}_{\rm mol} = (3^{+2}_{-1}) \cdot
      10^7 \, {\rm M_{\odot}}$.
\item The lack of a universal ${\cal N}_{\rm H_2}/I_{\rm ^{12}CO}$
      conversion constant and the presence of a mixture of thin and warm
      gas with cold and dense gas are likely to be general
      characteristics of the central regions of galaxies.
\end{enumerate}

\begin{acknowledgements}
We thank Leonardo Bronfman who has made available to us the $^{12}$CO
data of \cite{Bit87} in digital form before publication.  Christian
Henkel has kindly provided the LVG radiative transfer program.
RM was supported by a Heisenberg fellowship from the Deut\-sche
For\-schungs\-gemein\-schaft.
\end{acknowledgements}

\appendix
\section*{Appendix}

\section{Detailed Discussion of Figures}

\subsection{Spectral Line Shapes in Fig.~\protect\ref{VerglSpectra}} 
\label{LineShape} 

In Fig.~\ref{VerglSpectra}, spectra of both $^{12}$CO and C$^{18}$O
toward characteristic emission centers are shown: four examples of
single positions toward the Galactic bulge and two averaged over
several positions of Clump~2 \citep{Ban77}. In this section, we discuss
the presented spectra in detail particularly emphasizing the
applicability of the SCF to the molecular gas in the Galactic center
region. Besides comparing the spectra of the two CO isotopomers in
detail we make LTE estimates using the equations given in Section
14.8.1 of \cite{RoWi96}. For a detailed derivation in the case of
C$^{18}$O, in particular taking into account that all data refer to the
area of the telescope beam which is not infinitesimally small, see
\cite{Dah95}.

\subsubsection{Sgr\,A Region} 

In plot (a) of Fig.~\ref{VerglSpectra}, the \mbox{$^{12}$CO(1--0)} and
the \mbox{C$^{18}$O(1--0)} spectra toward $l = 0$\fdg{}0, $b =
0$\fdg{}0, near Sgr\,A, are shown.  Even though averaged over the
9\arcmin{}-telescope beam, there is considerable fine structure in the
spectra and the line shapes in $^{12}$CO and C$^{18}$O are very
different.  The two minima (B) and (D) in $^{12}$CO do not have
counterparts in C$^{18}$O. The minimum (D) coincides with the C$^{18}$O
maximum (a), the most intense feature in C$^{18}$O at this position.
Therefore, it is likely that the minima (B) and (D) in the $^{12}$CO
emission are caused by self absorption. Because of the rather narrow
line widths, this gas is likely to be local. Then, (A) and (C) show up
as maxima just because self absorption effects are minimal.

The so-called Expanding Molecular Ring \citep[EMR; see ]{Sco72,Kai+72}
is only marginally visible in \CO{18}, but is rather intense in
\CO{12}, especially at positive velocities.

At positive velocities, broad emission is present both in $^{12}$CO and
C$^{18}$O. The $^{12}$CO maximum (E) coincides roughly with maximum (b)
of C$^{18}$O. However, the most intense $^{12}$CO maximum (F) is only
present as an emission wing in C$^{18}$O. Because of the large velocity
extent, it is certain that these features arise in the Galactic center,
but the excitation of (E) and (F) must be different. If the SCF
(Eq.~\ref{Eq-H2-from-CO}) applies, the shapes of the $^{12}$CO and the
C$^{18}$O lines should be the same.  Because this is obviously not the
case, one or more of the requirements of this factor are not fulfilled
(see also Section~\ref{12COConvFac}). This alone is proof that the $J =
1 \to 0$ lines of $^{12}$CO and C$^{18}$O {\it cannot\/} both trace the
total H$_2$ mass, as it is commonly assumed for disk clouds.

\subsubsection{Sgr\,B2 Region} \label{LinShSgrB2}

In plot (b) of Fig.~\ref{VerglSpectra}, we present the spectra of the
\mbox{$^{12}$CO(1--0)} and the \mbox{C$^{18}$O(1--0)} transition toward
the position ($l = +0$\fdg{}6, $b = 0$\fdg{}0) near Sgr\,B2.  In
$^{12}$CO, six maxima can be recognized. Four of these are associated
with the main Galactic center emission and two with the EMR. In
general, the C$^{18}$O emission shows the same line shape as the
$^{12}$CO emission. The EMR peaks, which are the weakest features in
$^{12}$CO, are below our detection limit for C$^{18}$O. Thus, for this
position the SCF might be applicable.  However, compared to $^{12}$CO,
the peak components seem narrower and more pronounced over the extended
background in C$^{18}$O.

To check if the SCF is applicable, we evaluate the conversion factors
in detail for this position. The integrated intensities from $-$225.0
to +225.0~\kms{} are 25.3~K\,\kms{} for \mbox{C$^{18}$O(1--0)} and
1560~K\,\kms{} for \mbox{$^{12}$CO(1--0)}. From the SCF
(Eq.~\ref{Eq-H2-from-CO}, assuming \CO{12}/H$_2$ = 10$^{-4}$) one
finds the column density of $^{12}$CO as:
\begin{equation} 
\bar{\cal N}_{\rm ^{12}CO}({\rm SCF}) = 3.59 \cdot 10^{19} \ {\rm cm}^{-2}
\end{equation}

To derive the column density of C$^{18}$O, the excitation temperature
and beam-filling factor $f_{\rm beam}$ of CO must be estimated.  If the
SCF can be applied, the \mbox{$^{12}$CO(1--0)} line must be optically
thick.  Assuming that the excitation of the \mbox{CO(1--0)} transitions
is close to LTE and that the excitation temperature is the same for
both isotopomers, the relations from \cite{RoWi96}/\cite{Dah95} can be
used.

To apply these to our data, we need the beam filling factor,
$f_{\rm beam}$, both for C$^{18}$O and $^{12}$CO. For C$^{18}$O,
$f_{\rm beam}$ can be estimated by comparing our data with higher
angular resolution data.  \cite{Lin+95} presented
\mbox{C$^{18}$O(1--0)} data of the Sgr\,A region obtained with the
15\,m SEST telescope with a resolution of 45\arcsec. Their integrated
intensity toward the peak of the Sgr\,A complex is $\sim$\,50~K\,\kms{}
(on a \Tmb{}-scale).  Compared to our value of 17.3~K\,\kms, we
estimate $f_{\rm beam}$ for C$^{18}$O(1--0) to be:
\begin{equation} 
\label{BeamFillC18O} 
f_{\rm 9^{\prime}-beam}^{\rm C^{18}O} \approx 0.3 
\end{equation}
For the smoother and more widespread \mbox{$^{12}$CO(1--0)} emission
(see Fig.~\ref{12COContAllBit} and the \CO{12} presentation in
Paper~I), we estimate a beam-filling factor of:
\begin{equation} 
f_{\rm 9^{\prime}-beam}^{\rm ^{12}CO} \approx 0.5
\end{equation}

Using the assumptions mentioned above, the excitation temperature
\Tex{} is obtained from the \mbox{$^{12}$CO(1--0)} data. With \Tmb{} of 
\mbox{$^{12}$CO(1--0)} ranging from 2~K to 13.5~K, this results in a
\Tex{} from 7~K to 30~K, typically 20~K. For \mbox{C$^{18}$O(1--0)},
the emission is found to be optically thin, and \Tmb{} is typically
0.2~K. Then, the beam-averaged column density is:
\begin{equation}
\bar{\cal N}_{\rm C^{18}O}(T_{\rm ex} = 20\ {\rm K}) 
      = 2.65 \cdot 10^{16} \ {\rm cm}^{-2} 
\end{equation}

These column densities result in a $^{16}$O/$^{18}$O isotope ratio of
$\sim$\,1\,350, 2.7 times what is found in the local ISM and
in sharp contradiction to the value of 250 commonly adopted for the
Galactic center \citep{WiMa92,WiRo94}. 

To force agreement between the column densities of \CO{18} and
$^{12}$CO, one could revise the SCF or raise \Tex{} to 120\,K.  This
high a \Tex{} agrees with the \Tkin{} found for 34 Galactic center
cloud peaks \citep{Hut+93b} in NH$_3$.  LVG radiative transfer
calculations for the same cloud sample, applied to
\mbox{C$^{18}$O(1--0)} and \mbox{C$^{18}$O(2--1)}  lines
(\citeo{Hut93}, chapt.\ 5) give similar temperatures.  However, if
\Tex{} is this high, the beam averaged $^{12}$CO emission {\it
cannot\/} be optically thick, because then \Tmb{} should be $\sim
100$\,K and more which is clearly not the case. Thus, the SCF 
{\it cannot\/} be applied to $^{12}$CO.  On the other hand, if the 
$^{12}$CO emission {\it is\/} optically thick, \Tex{} can be {\it at 
most\/} 30~K. Then, the SCF for $^{12}$CO {\it does not\/} apply because 
the column density of $^{12}$CO is, even in the integration over the 
complete velocity range, too large by a factor of $\sim 5$ compared to 
C$^{18}$O.

In conclusion, even toward a position where, at first glance, it seems
that the SCF might be valid, it is shown that this is not the case.
This conclusion is even stronger if one considers other positions.

\subsubsection{Sgr\,C Region} 

Near Sgr\,C (plot (c) of Fig.~\ref{VerglSpectra}), the most prominent
feature in \mbox{$^{12}$CO(1--0)} both in peak intensity and line width
is the emission of the EMR. The main emission (B to C) is rather weak,
and, in addition, there is a narrow peak (A) which is probably of local
origin.

In C$^{18}$O, the features of the EMR are not visible. The main
Galactic center component is very similar to what is seen in $^{12}$CO
(b to c). The narrow peak (a) coincides almost exactly with the
respective $^{12}$CO peak (A), but is more pronounced.  The additional
(though tentative), narrow peak (d) corresponds to the 3-kpc-arm
\citep{RoOo60} emission and has no counterpart in $^{12}$CO; this is
probably superposed on the more intense main component. Such behaviour
is typical and is observed toward several positions.

These spectra confirm that the excitation conditions vary strongly with
velocity. In addition, the discrepancy between the column density of
$^{12}$CO, determined from the SCF, and the column density, determined
for the optically thin C$^{18}$O emission, is even larger than toward
Sgr\,B2.

\subsubsection{Sgr\,D Region}

Toward Sgr\,D, the $l=1$\fdg{}5-complex of \cite{Bal+88} (plot (d) of
Fig.~\ref{VerglSpectra}), the line profiles in \mbox{$^{12}$CO(1--0)}
and \mbox{C$^{18}$O(1--0)} agree very well.  The intense maxima (A/a)
coincide, and the different velocity extent (B to C in $^{12}$CO and b
to c in C$^{18}$O) can be explained with the lower signal-to-noise
ratio in the C$^{18}$O spectrum. This is also likely for the EMR peak
only visible in $^{12}$CO.  However, applying the same scheme as in
Appendix~\ref{LinShSgrB2}, it can again be shown that the SCF is not
applicable for this region.
 
Therefore, we will investigate the possibility whether such good
agreement in line shape can be explained by a {\it low\/} optical
depth, even for \mbox{$^{12}$CO(1--0)} emission. We find for the
intensities integrated over the velocity range from $-$225.0 to
+225.0~\kms{} 10.3~K\,\kms{} in case of \mbox{C$^{18}$O(1--0)} and
1\,270~K\,\kms{} in case of \mbox{$^{12}$CO(1--0)}. It is unlikely 
that the $^{12}$CO emission is truly optically thin. However, we can 
assume that the optical depth is of order unity, as is consistent with 
the nearly identical line shape. Then Eq.~\ref{Eq-Tex-tau-Tmb} holds, 
and the beam-averaged column density of $^{12}$CO is given by:
\begin{eqnarray}
\bar{\cal N}_{\rm ^{12}CO} & = & \frac{2.31 \cdot 10^{14}\,{\rm cm}^{-2}}
      {{\rm K\ km\ s}^{-1}} \cdot \frac{\tau_0^{\rm ^{12}CO(1-0)}}{1 - 
      e^{ - \tau_0^{\rm ^{12}CO(1-0)}}} \cdot \nonumber \\ 
  & & \frac{\displaystyle 
      \int T_{\rm MB}^{\rm ^{12}CO(1-0)}(v) \, {\rm d}v}
      {1 - \exp \left( - 5.53\,{\rm K} / T_{\rm ex} \right)}
      \label{Eq-tcol-12co-thin} 
\end{eqnarray}
Adopting the standard $^{16}$O/$^{18}$O ratio in the Galactic center,
one can set this expression equal to 250 times the column density of
the optically thin \CO{18} and solve for the optical depth:
\begin{eqnarray} 
\frac{\tau_0^{\rm ^{12}CO(1-0)}} 
      {1 - e^{ - \tau_0^{\rm ^{12}CO(1-0)}}} & = & 
      \frac{2.43 \, \left[  1 - \exp \left( - 5.53\,{\rm K} / T_{\rm ex} 
      \right) \right] } {2.31 \, \left[ 1 - \exp \left( - 5.27\,{\rm K} 
      / T_{\rm ex} \right) \right] } \cdot \nonumber \\ 
  & & \frac{\displaystyle 250 \, \int T_{\rm MB}^{\rm C^{18}O(1-0)}(v) 
      \, {\rm d}v} {\displaystyle \int T_{\rm MB}^{\rm ^{12}CO(1-0)}(v) 
      \, {\rm d}v}
      \label{Eq-tau12-RatInt} 
\end{eqnarray}
Note that $f_{\rm beam}$ plays no role as long as the beam and cloud
sizes are comparable. The dependence on \Tex{} is weak and the fraction
containing it is equal to  $\sim$\,1.1. Therefore, the optical depth of
the $^{12}$CO emission can be determined from the integrated
intensities and one obtains:
\begin{equation}
\tau_0^{\rm ^{12}CO(1-0)}(l=1.\!\!^{\circ}2,\,b=0.\!\!^{\circ}0) 
\approx 1.9
\end{equation}
Thus, surprisingly, moderate optical depths are obtained from an
analysis of the data, even for one of the most intense $^{12}$CO peaks
in the Galactic center region.

\subsubsection{Northern Clump~2 Region}

From Figs.~\ref{C18OContChan} and \ref{12COContChan} in
\pageref{12COContChan}, we noted that the velocity structure of the
northern and the southern region of Clump~2 might be completely
different. A detailed analysis of the spectra shows that this is indeed
the case. Therefore, we averaged Clump~2 spectra separately for the
northern and the southern part.  The positions included in the averages
were chosen for both parts from the presence of at least some C$^{18}$O
emission and of similar velocity structure, as suggested by
Fig.~\ref{C18OContChan} in \pageref{C18OContChan}.

In plot (e) of Fig.~\ref{VerglSpectra}, we show the average of 14
positions toward the northern region of Clump~2.  In $^{12}$CO, this
northern Clump~2 region shows a broad plateau of emission (B to C) from
about $-$30 to +180~\kms. It has a nearly constant intensity of some
2.5~K, except for the strong peak (A) at $\sim$\,+20~\kms with an
intensity of about 4.3~K. Two weak side peaks (D and E) at $\sim\,-$70
and $\sim\,-$45~\kms{} can be recognized. Peak (E) probably belongs to
the 3-kpc-arm because its velocity matches the expected velocity of
this feature at this Galactic longitude \citep{Ban77,Ban80,Ban86}.

In C$^{18}$O, the emission is very weak, indicating (as discussed)
moderate to low optical depth in $^{12}$CO emission. Thus, the total
molecular mass of Clump~2 will be significantly less than suggested by
the SCF. The only feature clearly visible in C$^{18}$O is the peak (a)
at $\sim$\,+20~\kms{} with wings (b to c) from about +5 to +60~\kms,
similar to peak (A) and its wings visible in $^{12}$CO.  Other features
visible in $^{12}$CO are obviously too weak to show up in C$^{18}$O.

\subsubsection{Southern Clump~2 Region}

In plot (f) of Fig.~\ref{VerglSpectra}, we show the average of 8
positions toward the southern region of Clump~2.

In the $^{12}$CO spectrum of this region of Clump~2, 5 maxima can be
clearly distinguished. The two more intense ones (C and E) are at about
+5 and +95~\kms; the weaker features are at about $-$40, $-$25, and
+45~\kms. The EMR tends to be found at more negative latitudes
and more positive velocities toward more positive longitudes. In this
plot, it appears with a peak at about +225~\kms. Note the weak emission
which connects the EMR tail with the main emission components and the
weak extended wing toward velocities as negative as $-$110~\kms.

In C$^{18}$O, peak (b) at about +105~\kms{} is the only clearly visible
feature. It roughly coincides with the $^{12}$CO peak (E) although the
maximum is at higher velocities. Another peak, (a), may be present at
about +20~\kms. However, around about $-$5~\kms, the spectrum seems to
show some artifacts. In addition, there is no counterpart peak visible
in $^{12}$CO. Thus, this emission may be spurious. However, this peak
appears at the same velocity as peak (a) in the northern part of
Clump~2. Because we have already found lacks of agreement between
$^{12}$CO and C$^{18}$O, this feature could indicate a component
present over the entire Clump~2.

\subsection{The Features in Fig.~\protect\ref{RatioContChan}} 
\label{FeatFigRatCoCh} 

In Fig.~\ref{RatioContChan}, the ratio of the intensity of
\mbox{$^{12}$CO(1--0)} to \mbox{C$^{18}$O(1--0)} emission in the
Galactic center region, integrated over velocity intervals of 50~\kms,
is displayed. Here we give a detailed description of the plots.

At $-$150~\kms{} and at $-$100~\kms, the Sgr\,C region is most
prominent. It shows high ratios reaching $>$\,75, peaking with a ratio
of $\sim$\,90 at ($l = -0$\fdg{}47, $b = +0$\fdg{}13) and ($l =
-0$\fdg{}6, $b = 0$\fdg{}0), respectively. Most of the data points are
lower limits. However, the ratio of the maximum of the Sgr\,C region at
$-$100~\kms{} is an actual value, not merely a limit.

At $-$50~\kms{} (the 3-kpc-arm) and 0~\kms{} (the widespread component
of the Galactic center, but also local gas), no very high ratios are
found.  However, the ratios show much structure, and typically range
from 30 to 80 in the areas where C$^{18}$O is above the detection limit
(refer to Fig.~\ref{C18OContChan} in \pageref{C18OContChan}). Note,
that at $-$50~\kms, the Sgr\,A region and the region at ($l =
+0$\fdg{}9, $b = 0$\fdg{}0) appear as valley-like minima with ratios
$<$\,40. These are not lower limits and are surrounded by rims of
higher ratios. At 0~\kms, the same effect can be seen again in the
Sgr\,A region as well as in the Sgr\,B2 and in the region at ($l =
+1$\fdg{}05, $b = -0$\fdg{}15).

In the velocity ranges centered at +50~\kms{} and at +100~\kms, the
ratio reaches values of 90 to 120 in large areas, and in a few
positions these are higher still, reaching values $>$\,200.  The most
prominent areas are the Sgr\,D region \citep[the $l=1$\fdg{}5-complex
of ]{Bal+88} and Clump~2. In the Sgr\,D region, ratios of 90 to 120 are
reached at +50~\kms{} in the area from $l = +1$\fdg{}1 to +1\fdg{}8 and
from $b = -0$\fdg{}35 to +0\fdg{}2 peaking at ($l = +1$\fdg{}4, $b =
-0$\fdg{}15) with a ratio of $>$\,200 and at +100~\kms{} in the area
from $l = +0$\fdg{}9 to +1\fdg{}6 and from $b = -0$\fdg{}3 to
+0\fdg{}45 peaking at ($l = +1$\fdg{}5, $b = +0$\fdg{}3) also with a
ratio of $>$\,200.  In Clump~2 at +50~\kms, the ratio reaches such high
values in the area from $l = +3$\fdg{}0 to +3\fdg{}35 and from $b =
+0$\fdg{}3 to +0\fdg{}65 peaking at ($l = +3$\fdg{}3, $b = +0$\fdg{}5)
with a ratio of nearly 170. At +100~\kms, such high ratios are achieved
in the area from $l = +2$\fdg{}9 to +3\fdg{}35 and from $b =
-0$\fdg{}05 to +0\fdg{}65 peaking at ($l = +3$\fdg{}15, $b =
+0$\fdg{}15) with a ratio of $\sim$\,160. The absolutely highest ratio
over all velocities is the maximum of the Sgr\,D region at +100~\kms.
In addition, there are 2 definite valley-like minima, where the
determined ratios are not lower limits.  These minima are the Sgr\,B2
region at +50~\kms{} (ratio $\losim 40$) and the region around ($l =
+1$\fdg{}05, $b = 0$\fdg{}0) at +100~\kms{} (ratio $\sim$\,70).  Both
regions coincide with maxima in the C$^{18}$O emission and are
surrounded by areas with higher ratios.

At +150~\kms{} and even more at +200~\kms, almost all ratios are lower
limits.  However, at +150~\kms{} there are two regions where even the
lower limit of the ratio exceeds 90, namely toward ($l = -0$\fdg{}75,
$b = +0$\fdg{}05) near the Sgr\,C region and toward ($l = +0$\fdg{}45,
$b = +0$\fdg{}3). Toward three further regions, the ratio reaches
values of about 75, i.e.\ toward ($l = +0$\fdg{}15, $b = +0$\fdg{}15)
near the Sgr\,A region toward ($l = +1$\fdg{}35, $b = +0$\fdg{}4) near
the Sgr\,D region and toward ($l = +1$\fdg{}2, $b = -0$\fdg{}15). Most
prominent at +150~\kms{} is Clump~2 where ratios of 90 to 120 are found
in the area $l = +2$\fdg{}95 to +3\fdg{}35 and from $b = 0$\fdg{}0 to
+0\fdg{}55 peaking at ($l = +3$\fdg{}15, $b = +0$\fdg{}15) with a ratio
of about 135 where the ratio is a real value and not a lower limit.
Valley-like minima surrounded by areas with higher ratios are present
at ($l = +1$\fdg{}0, $b = +0$\fdg{}1, ratio $\sim$\,20) close to the
minimum at +100~\kms{} and at ($l = +1$\fdg{}5, $b = 0$\fdg{}0) where
even no $^{12}$CO emission is found.

\end{document}